\documentclass[11pt,a4paper]{elsarticle}
\pdfoutput=1
\usepackage[margin=3.6cm]{geometry}
\usepackage{float}
\usepackage{subfigure}
\usepackage{braket}
\usepackage{amsmath}
\usepackage{amssymb}
\biboptions{sort&compress}
\title{Cosmic axion background propagation in galaxies}
\author[a]{Francesca V. Day}
\address[a]{Rudolf Peierls Centre for Theoretical Physics, University of Oxford,\\
1 Keble Road, Oxford, OX1 3NP, United Kingdom}
\ead{francesca.day@physics.ox.ac.uk}

\newcommand{\be}{\begin{equation}}
\newcommand{\ee}{\end{equation}}
\newcommand{\bea}{\begin{eqnarray}}
\newcommand{\eea}{\end{eqnarray}}

\newcommand{\half}{\frac{1}{2}}

\begin{document}

\begin{abstract}

Many extensions of the Standard Model include axions or axion-like particles (ALPs). Here we study ALP to photon conversion in the magnetic field of the Milky Way and starburst galaxies. By modelling the effects of the coherent and random magnetic fields, the warm ionized medium and the warm neutral medium on the conversion process, we simulate maps of the conversion probability across the sky for a range of ALP energies. In particular, we consider a diffuse cosmic ALP background (CAB) analogous to the CMB, whose existence is suggested by string models of inflation. ALP-photon conversion of a CAB in the magnetic fields of galaxy clusters has been proposed as an explanation of the cluster soft X-ray excess. We therefore study the phenomenology and expected photon signal of CAB propagation in the Milky Way. We find that, for the CAB parameters required to explain the cluster soft X-ray excess, the photon flux from ALP-photon conversion in the Milky Way would be unobservably small. The ALP-photon conversion probability in galaxy clusters is 3 orders of magnitude higher than that in the Milky Way. Furthermore, the morphology of the unresolved cosmic X-ray background is incompatible with a significant component from ALP-photon conversion. We also consider ALP-photon conversion in starburst galaxies, which host much higher magnetic fields. By considering the clumpy structure of the galactic plasma, we find that conversion probabilities comparable to those in clusters may be possible in starburst galaxies.

\end{abstract}


\maketitle
\flushbottom

\section{Introduction}

Axions and axion-like particles (ALPs) arise in many extensions of the Standard Model as pseudo-Nambu-Goldstone bosons of broken symmetries. A generic ALP is an ultra-light pseudo-scalar singlet under the Standard Model gauge group. Throughout this work we will consider massless ALPs with no coupling to QCD. In the low energy effective field theory, an explicit mass term in the Lagrangian is forbidden by a shift symmetry $a(x) \to a(x) + {\rm constant}$. We expect non-renormalizable couplings between the ALP and the Standard Model suppressed by the high scale $M$. In this work we explore the phenomenology of the dimension five $a \gamma \gamma$ coupling. In addition to the Standard Model Lagrangian, we therefore have:

\begin{equation}
\label{Lagrangian}
\mathcal{L}_{{\rm ALP}} =  \frac{1}{2} \partial_{\mu} a \partial^{\mu} a + \frac{1}{4M} a F \tilde{F} = \frac{1}{2} \partial_{\mu} a \partial^{\mu} a + \frac{a}{M} {\bf E} \cdot {\bf B}.
\end{equation}
The term $\mathcal{L} \supset \frac{a}{M} {\bf E} \cdot {\bf B}$ leads to ALP-photon interconversion in the presence of a background magnetic field. The mass scale $M$ is model dependent and so is a priori undetermined. Empirical limits on $M$ may be derived from astrophysical observations and from axion search experiments, as reviewed in \cite{ALPquest}. For low mass ALPs ($m_{a} \lesssim 10^{-10} \, {\rm eV}$), the strongest bounds on $M$ arise from observations of the SN1987a supernova in the Large Magellanic Cloud. In ALP extensions of the Standard Model, we would expect an ALP burst coincident with the neutrino burst. This ALP burst would be observable as a gamma ray flux following ALP-photon conversion in the Milky Way magnetic field. The non-observation of such a gamma ray flux leads to the bound $M \gtrsim 2 \times 10^{11} \, {\rm GeV}$ \cite{SN1987A1,SN1987A2,SN1987A3}.\\

\begin{figure} [h]
\includegraphics[scale=0.7]{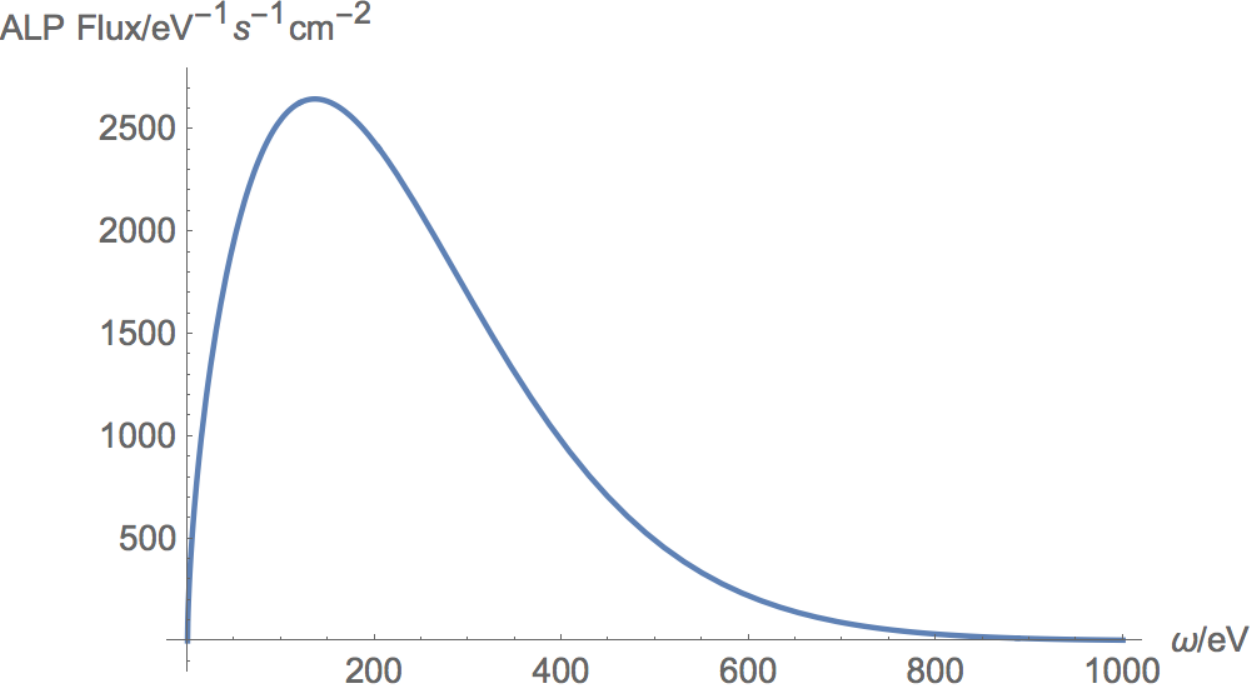}
\label{ALPflux}
\caption{The predicted ALP flux $\frac{d \Phi_{a}}{dE}$ for $E_{{\rm CAB}} = 200 \, {\rm eV}$ and $\Delta N_{{\rm eff}} = 0.5$}
\end{figure}

A primordially generated, thermally produced cosmic ALP background (CAB), analogous to the CMB, is a natural prediction of string theory models of inflation \cite{cosmopheno}. The CAB has a quasi-thermal energy spectrum that is red shifted to soft X-ray energies today. The constituent ALPs act as dark radiation - extra relativistic degrees of freedom conventionally parametrised by the equivalent excess in the number of neutrino species $\Delta N_{{\rm eff}}$. Current measurements of $\Delta N_{{\rm eff}}$ are consistent both with zero and with a significant dark radiation component \cite{Planck}. The ALP number density in the CAB between energies $E$ and $E + dE$ is:

\be 
dN \left( E \right) = A X\left( E \right) dE,
\ee
where $X\left( E \right)$ is the shape of the CAB energy spectrum and A its normalisation. The spectral shape is predicted by the general string inflation scenario described in \cite{cosmopheno}, and may be found by numerically solving the Friedmann equations for ALP production and redshift. The resulting spectrum is fit well by the function

\be
\label{X}
X\left( E \right) =  E^{q} e^{ -a E^{r}}.
\ee
The constants $q$, $a$ and $r$ are found by fitting equation \eqref{X} to a numerical solution of the equations of motion, and in general depend on the mean ALP energy $E_{{\rm CAB}}$. In a typical string inflation model, $E_{{\rm CAB}} \sim \mathcal{O} (100 \, {\rm eV})$. The overall normalisation of the spectrum is model dependent but may be measured by its contribution to $\Delta N_{\rm{eff}}$. We will therefore find the normalisation constant $A$ by setting the CAB contribution to $\Delta N_{\rm{eff}}$. This is related to the CAB energy density by:

\be 
\rho_{\rm{CAB}} = \Delta N_{\rm{eff}} \frac{7}{8} \left( \frac{4}{11} \right)^\frac{4}{3} \rho_{\rm{CMB}}.
\ee
The flux $d \Phi_{a} \left( E \right)$ of ALPs with energies between $E$ and $E + dE$ is then:

\begin{equation}
d \Phi_{a} \left( E \right)  = dN \left( E \right) \frac{c}{4},
\end{equation}
so,
\begin{equation}
\frac{d \Phi_{a}}{dE} = AX(E) \frac{c}{4}.
\end{equation}
The predicted spectrum of the CAB background for $E_{{\rm CAB}} = 200 \, {\rm eV}$  and $\Delta N_{{\rm eff}} = 0.5$ is shown in figure \ref{ALPflux}. In this case, the parameters in equation \eqref{X} are found to be $q = 0.62$, $r = 1.5$, $a = 2.6 \times 10^{-4} \, {\rm eV}^{-1.5}$.

For sufficiently large $\frac{1}{M}$ and CAB flux, ALP-photon conversion offers the possibility of detecting a CAB as an excess of soft X-ray photons from environments with a sufficiently strong and coherent magnetic field \cite{13053603}.  A natural place to search for this effect is in galaxy clusters, which host $1 - 10 \, \mu {\rm G}$ fields over Mpc distances. Furthermore, there is a long standing excess in the soft X-ray ($E \lesssim 400 \, {\rm eV}$) flux observed from galaxy clusters, above the predicted thermal emission from the intra-cluster medium. It was suggested in \cite{13053603} that CAB to photon conversion in galaxy clusters could be the source of this soft X-ray excess. Detailed simulations of this process have been carried out for the Coma \cite{13123947, outskirts}, A665, A2199 and A2255 \cite{14114172} galaxy clusters. These show that CAB to photon conversion can consistently explain the observed excess in Coma, A2199 and A2255 as well as the non-observation of an excess in A665, within astrophysical uncertainties.

While by no means conclusive, this hint of new physics motivates studying the consequences of a CAB in other astrophysical systems. Galaxies also host magnetic fields and are therefore potential ALP to photon converters, as discussed in \cite{SN1987A1,Simet,12070776,13021208,Wouters,Fairbairn,14047741,14101867,SN1987A3}. Note that we come to qualitatively different conclusions than those in \cite{Fairbairn}. This is discussed further in the Appendix. ALPs from a CAB may convert to X-ray photons in the Milky Way. This would contribute to the unresolved cosmic X-ray background - the diffuse X-ray intensity observed across the sky after subtracting the integrated emission from all detected point sources. Within standard physics, the unresolved cosmic X-ray background could arise from the Local Bubble and the warm-hot intergalactic medium \cite{CDF,SXB}. There is also room for more exotic contributions, such as decaying dark matter or the CAB considered here. However, unlike in the case of the galaxy cluster soft X-ray excess, there is no problem explaining the cosmic X-ray background within the framework of standard physics. The possibility that the cosmic X-ray background is related to conversion of a CAB to photons in the Milky Way's magnetic field was first considered in \cite{13053603,Fairbairn}.

The magnetic field in a starburst galaxy (a galaxy with a very high rate of star formation) is typically an order of magnitude higher than that in the Milky Way, suggesting a substantially higher rate of ALP-photon conversion. We therefore also estimate the ALP to photon conversion probability in starburst galaxies. In both cases, the warm ionized and neutral gas in the galaxy also plays a significant role in determining the ALP to photon conversion probability, as discussed in sections 2 and 5.

In this paper, we will discuss the phenomenology and potential observational consequences of a CAB's passage through the Milky Way. We will also consider ALP to photon conversion in the high magnetic field, high plasma density environment of starburst galaxies. In section 2 we consider in more detail the propagation of the ALP-photon vector in galactic environments, focusing on the effect of the magnetic field, the warm ionised medium and the warm neutral medium. In section 3, we describe our model of the Milky Way environment. In section 4, we present and discuss our results for ALP-photon conversion in the Milky Way. In section 5, we discuss some caveats and additional relevant effects. In particular we derive the conditions under which the clumpiness of the warm ionized gas in galaxies becomes relevant for ALP-photon conversion. In section 6 we apply this to estimate the ALP-photon conversion probability in starburst galaxies. We conclude in section 7.

\section{ALP-photon conversion}
\label{conversion}

The ALP-photon coupling is suppressed by an energy scale $M$ much larger than the physical energies involved. It is therefore sufficient to simulate ALP-photon conversion using the classical equation of motion derived from \eqref{Lagrangian}, and neglecting higher dimension terms. We further assume that the ALP wavelength is much shorter than the scale over which its environment changes, allowing us to linearise the equations of motion. This condition is abundantly satisfied for X-ray energy ALPs in astrophysical environments. The ALP-photon equations of motion are conveniently described by combining the ALP with the two photon polarizations in an ALP-photon vector. The linearised equation of motion for an ALP-photon vector of energy $\omega$ is then:

\begin{equation}
\label{propfree}
\left( \omega + \left( \begin{array}{ccc}
\Delta_{\gamma} & \Delta_{F} & \Delta_{\gamma a x} \\
\Delta_{F} & \Delta_{\gamma} & \Delta_{\gamma a y} \\
\Delta_{\gamma a x}  & \Delta_{\gamma a y} & 0 \end{array} \right)
 - i\partial_{z} \right) \left( \begin{array}{c}
\mid \gamma_{x} \rangle\\
\mid \gamma_{y} \rangle\\
\mid a \rangle \end{array} \right) = 0.
\end{equation}
ALP-photon mixing is controlled by $\Delta_{\gamma a i} = \frac{B_{i}}{2M}$ where $i = x,y$ are the two directions perpendicular to the direction of travel. $\Delta_{F}$ describes Faraday rotation between the two photon polarizations. This effect is not relevant to ALP-photon conversion, and so we set $\Delta_{F} = 0$. The photon components are given an effective mass by their interactions with free electrons in the surrounding medium. This effective photon mass is equal to the plasma frequency - the frequency of charge density oscillations in the surrounding plasma. This is given by $\omega_{pl} = \left( 4 \pi \alpha \frac{n_{e}}{m_{e}} \right) ^ {\frac{1}{2}}$, where $n_{e}$ is the free electron density. We then have :

\be
\label{Deltagamma}
\Delta_{\gamma} = \frac{-\omega_{pl}^{2}} {2 \omega} =  -\frac{4 \pi \alpha n_{e}}{2 \omega m_{e}}.
\ee
As we do not measure the photon polarization, we simply add the conversion probabilities for each polarization. For an initially pure ALP state, in our semi-classical approximation the conversion probability after a distance $L$ is:

\be 
P_{a \to \gamma}(L) = |\braket{1,0,0|f(L)}|^{2} +  |\braket{0,1,0|f(L)}|^{2},
\ee
where $\ket{f(L)}$ is the final state after a distance $L$ as determined by equation \eqref{propfree}. The ALP to photon conversion probability $P_{a \to \gamma}$ is proportional to $\frac{B_{x}^{2} + B_{y}^{2}}{M^{2}}$ in the limit $\frac{B}{M} \ll 1$. A non-zero electron density in the propagation environment gives an effective mass to the photon, causing decoherence between the ALP and photon components and hence suppressing $P_{a \to \gamma}$. 

For constant electron density and magnetic field, there is an analytic solution for the conversion probability. We identify two angles associated with the propagation:
\bea
{\rm tan} \left( 2 \theta \right) & = & 2.8 \times 10^{-3} \times \left( \frac{10^{-3} \, {\rm cm}^{-3}}{n_{e}} \right) \left( \frac{B_{\perp}}{1 \, \mu {\rm G}} \right) \left( \frac{\omega}{1 \, {\rm keV}} \right) \left( \frac{10^{13} \, {\rm GeV}}{M} \right), \\
\Delta & = & 0.053 \times \left( \frac{n_{e}}{10^{-3} \, {\rm cm}^{-3}} \right) \left(\frac{1 \, {\rm keV}}{\omega} \right) \left( \frac{L}{1 \, {\rm kpc}} \right).
\eea
For a single domain of length $L$, the conversion probability is then
\be
\label{singleDomain}
P(a\ \to \gamma) = \sin^2 \left( 2 \theta \right) \sin^2 \left( \frac{\Delta}{\cos 2 \theta} \right).
\ee
In a more general case, for $P_{a \to \gamma} \ll 1$, we find:

\be
\label{semi-analytic}
P_{a\to \gamma}(L)=\sum_{i=x,y}\left|\int^{L}_{0} dz e^{i \varphi(z )} \Delta_{\gamma a i}(z)\right|^2 \, ,
\ee
where,
\be
\label{phi}
\varphi(z)  = \int^{z}_{0} dz' \Delta_{\gamma}(z') = - \frac{1}{2\omega} \int^{z}_{0} dz' \omega_{pl}^2(z') \, .
\ee
As shown in equation \eqref{Deltagamma}, $\Delta_{\gamma}(z) \propto n_{e}$, and so the electron density has the effect of rotating the probability amplitudes $\braket{1,0,0|f(L)}$ and $\braket{0,1,0|f(L)}$ in the complex plane as L increases, suppressing the efficacy of the magnetic field in increasing the conversion probability over increasing distances.\\

The Milky Way is almost opaque to low energy X-rays due to photoelectric absorption from the warm neutral medium. We capture this effect in our equation of motion for the ALP-photon vector by including a damping parameter $\Gamma(z)$ that describes the attenuation of the photon components. The new equation of motion no longer describes a closed quantum system - the Hamiltonian for the ALP-photon vector alone is no longer Hermitian. We therefore use a density matrix formalism:

\begin{equation}
\label{H}
H = \left( \begin{array}{ccc}
\Delta_{\gamma} & 0 & \Delta_{\gamma a x} \\
0 & \Delta_{\gamma} & \Delta_{\gamma a y} \\
\Delta_{\gamma a x}  & \Delta_{\gamma a y} & 0 \end{array} \right)
 -
 \left( \begin{array}{ccc}
i \frac{\Gamma}{2} & 0 & 0 \\
0 & i \frac{\Gamma}{2} & 0 \\
0  & 0 & 0 \end{array} \right),
\end{equation}
\begin{equation}
\rho =  \left( \begin{array}{c}
\mid \gamma_{x} \rangle\\
\mid \gamma_{y} \rangle\\
\mid a \rangle \end{array} \right) \otimes
\left( \begin{array}{ccc}
\mid \gamma_{x} \rangle & \mid \gamma_{y} \rangle & \mid a \rangle \end{array} \right) ^{*}  \\ ,
\end{equation}
\begin{equation}
\rho(z) = e^{-iHz} \rho(0) e^{iH^{\dagger}z} .
\end{equation}
To simulate the conversion probabilities, we discretize each line of sight into domains of length $\delta z$:
\begin{equation}
\label{discrete}
\rho_{k} = e^{-iH_{k} \delta z} \rho_{k - 1} e^{iH^{\dagger}_{k} \delta z},
\end{equation}
where $\rho_{k}$ is the density matrix in the $k$th domain and $H_{k}$ is the Hamiltonian defined using the magnetic field, electron density and neutral hydrogen density in the centre of the $k$th domain. 

\section{The Milky Way environment}

Three properties of the Milky Way's interstellar medium are relevant to ALP-photon conversion - the magnetic field, the free electron density provided by the warm ionized medium and the opacity to X-rays provided by the warm neutral medium. In this section, we describe our model for each of these components. We leave a discussion of various caveats to and justifications of this model, in particular the clumpiness of the electron density, to section 5.

\subsection{Magnetic field}
\begin{figure} [h]
\includegraphics[scale=0.7]{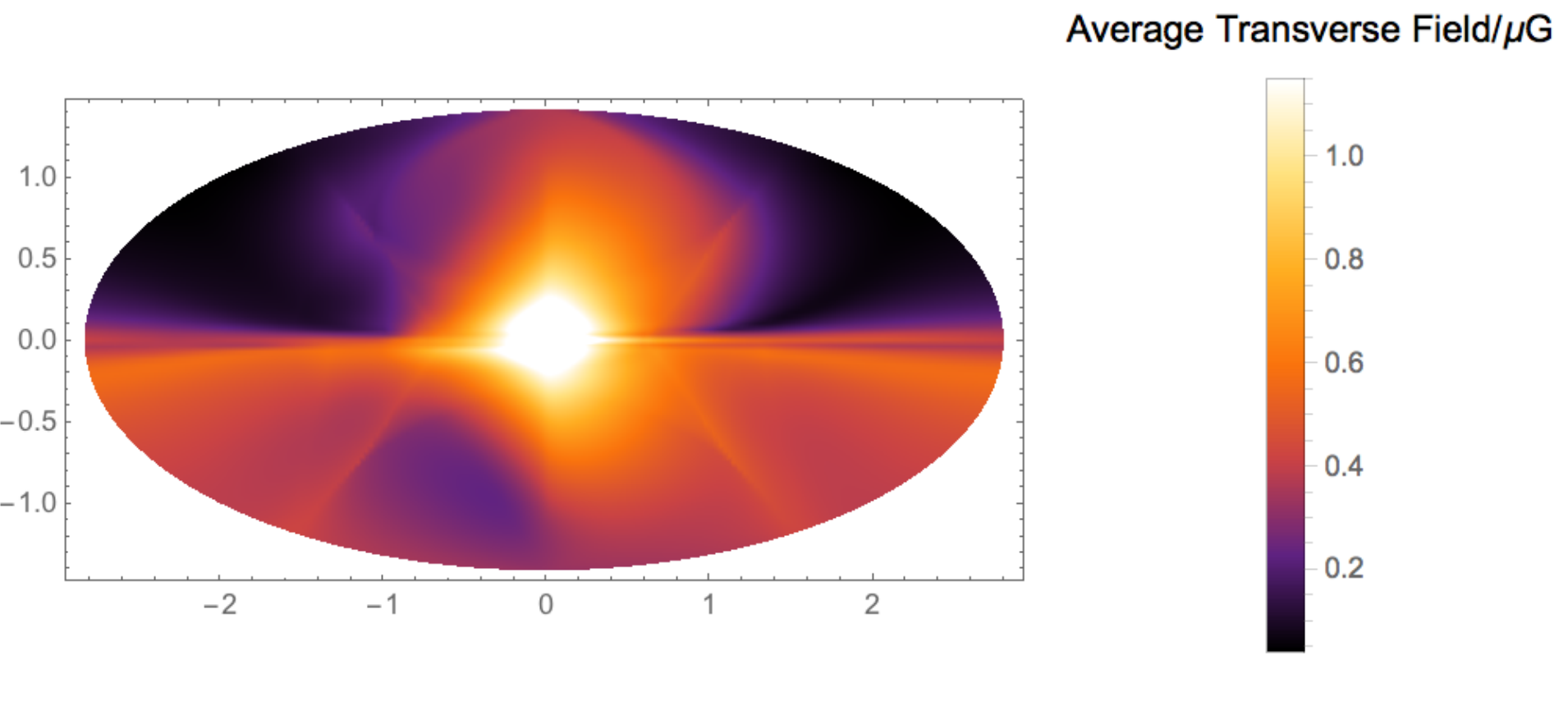}
\caption{The average coherent transverse magnetic field for lines of sight starting 20 kpc from the Earth. Galactic latitude increases vertically in the plot, while galactic longitude increases to the right. The centre of the plot corresponds to a line of sight in the direction of the galactic centre.}
\label{field}
\end{figure}

We use the recent model by Jansson and Farrar \cite{Farrar1, Farrar2}, based on 40,000 extra-galactic Faraday rotation measures. The magnetic field is the sum of three components - the coherent field, the random field and the striated field. The coherent field has large scale structure on the scale of the Milky Way with typical field strengths of a few $\mu {\rm G}$. The coherent field is modelled as the sum of a disc field, which follows the spiral arms of the Milky Way; a halo field above and below the disc; and an `X field' which points out of the plane of the Milky Way. The radial extent of the halo field is much greater in the South of the galaxy than in the North. The coherent field model in \cite{Farrar1} artificially excludes the central 1 kpc of the Milky Way. We therefore augment the model with a 5 $\mu {\rm G}$ radially constant poloidal field with vertical scale height 1 kpc in the central 1 kpc only. A full sky map of the average coherent field is shown in figure \ref{field}.

The random field has a set magnitude with a disc and halo component, but its direction is randomized with a coherence length of $\mathcal{O} \left(100 \, \rm{pc} \right)$, the typical size of a supernova outflow. The magnitude of the random field is typically a few times higher than that of the coherent field. The striated field has a magnitude 1.2 times that of the coherent field with its sign randomized on coherence scales of $\mathcal{O} \left(100 \, \rm{pc} \right)$. We see from equation \eqref{semi-analytic} that the conversion probability increases with the coherence length of the magnetic field. Indeed, for the majority of the Milky Way the coherent field gives the dominant contribution to ALP-photon conversion. The exception to this is in the disc of the Milky Way, where the random field is $\mathcal{O} \left( 10 \, \mu {\rm G} \right)$ whereas the coherent field is $\mathcal{O} \left( 1 \, \mu {\rm G} \right)$. Additionally, the coherent field often reverses sign between the spiral arms, decreasing its coherence length in the disc. We therefore use all three field components in modelling ALP-photon conversion in the Milky Way. The random and striated fields are implemented with respect to each line of sight - the direction and sign respectively are randomised every 100 pc along each ALP-photon path separately. This simple implementation clearly does not give a realistic picture of the random and striated fields across the Milky Way, but is adequate for modelling their effects on ALP-photon conversion.

\subsection{Electron density}

As described above, the photon gains an effective mass through interactions with surrounding free electrons. This mass suppresses ALP-photon conversion, as shown in equations \eqref{Deltagamma} and \eqref{semi-analytic}. We use the thin and thick disc components of the NE2001 \cite{NE2001} model of the Milky Way electron density:

\begin{equation}
\begin{aligned} 
g_{{\rm thick}} \left( r \right) &= \frac{{\rm cos} \left( \frac{\pi r}{34 \, {\rm kpc}}\right)}{{\rm cos} \left( \frac{\pi R_{\odot} }{34 \, {\rm kpc}}\right)} H\left(17 \, {\rm kpc} - r \right), \\
g_{{\rm thin}} \left( r \right) &= {\rm e}^{-\left( \frac{r - 3.7 \, {\rm kpc}}{1.8 \, {\rm kpc}} \right)^{2}}, \\
n_{{\rm thick}} \left( r,z \right) &= 0.035 \, {\rm cm}^{-3} g_{{\rm thick}} \left( r \right) {\rm sech}^{2} \left( \frac{z}{0.95 \, {\rm kpc}} \right), \\
n_{{\rm thin}} \left( r,z \right) &= 0.09 \, {\rm cm}^{-3} g_{{\rm thin}} \left( r \right) {\rm sech}^{2} \left( \frac{z}{0.14 \, {\rm kpc}} \right), \\
n_{e} \left( r,z \right) &= n_{{\rm thick}} \left( r,z \right) +n_{{\rm thin}} \left( r,z \right),
\end{aligned}
\end{equation}
where $\left( r, z \right)$ are cylindrical polar coordinates centred at the galactic centre, $R_{\odot} = 8.5 \, {\rm kpc}$ is the distance to the Sun and $H \left( x \right)$ is the Heaviside step function. This model predicts unphysically low electron densities at large radii. While this is not important for many astrophysical phenomena, which depend only on line of sight integrals of $n_{e}$, it can have a large effect on $P_{a \to \gamma}$. We therefore enforce a minimum electron density of $n_{{\rm min}} = 10^{-7} \, {\rm cm}^{-3}$, approximately the electron density of inter-galactic space.

\subsection{Photoelectric absorption by the warm neutral medium}
\label{absorption}

As explained in section 2, we model photoelectric absorption with the damping parameter $\Gamma$, which describes the attenuation of the photon component of the ALP-photon vector. This is conventionally parameterzied by the effective cross section with respect to neutral hydrogen, so that $\Gamma \left( \omega \right) = \sigma_{{\rm eff}} \left( \omega \right) \left( n_{HI} + 2n_{H_{2}} \right)$, where $n_{HI} + 2n_{H_2}$ is the density of neutral hydrogen. ($HI$ refers to atomic hydrogen and $H_{2}$ to molecular hydrogen.) Photoelectric absorption by heavier elements (which is dominant for $\omega \gtrsim 1 \, {\rm keV}$) is included in the effective cross section $\sigma_{{\rm eff}} \left( \omega \right)$ by assuming solar abundances for the relative densities of hydrogen and heavier elements. We use effective cross section values from \cite{crosssections} - we note in particular that $\sigma_{{\rm eff}} \left( \omega \right)$ is highly energy dependent, ranging from $\sigma_{{\rm eff}} \left( 100 \, {\rm eV} \right) = 5.7 \times 10^{-20} \, {\rm cm}^{2}$ to $\sigma_{{\rm eff}} \left( 2 \, {\rm keV} \right) = 4.3 \times 10^{-23} \, {\rm cm}^{2}$. We use the neutral hydrogen densities given in \cite{H}:

\begin{equation}
n_{HI} =
\begin{cases}
    0.32 \, {\rm cm}^{-3} {\rm exp} \left( - \frac{r}{18.24 \, {\rm kpc}} - \frac{\shortmid z \shortmid} {0.52 \, {\rm kpc}} \right),& \text{if } r \geq 2.75 \, {\rm kpc}\\
    0,              & \text{otherwise}
\end{cases}
\end{equation}
\begin{equation}
n_{H2} = 4.06 \, {\rm cm}^{-3} {\rm exp} \left( - \frac{r}{2.57 \, {\rm kpc}} - \frac{\shortmid z \shortmid} {0.08 \, {\rm kpc}} \right).
\end{equation}

\section{The Milky Way: Results and Discussion}

We now apply equations \eqref{H} and \eqref{discrete} to the ALP-photon conversion in the Milky Way, as described above.

\subsection{Conversion Probabilities}

\begin{figure} [h]
\includegraphics[scale=0.7]{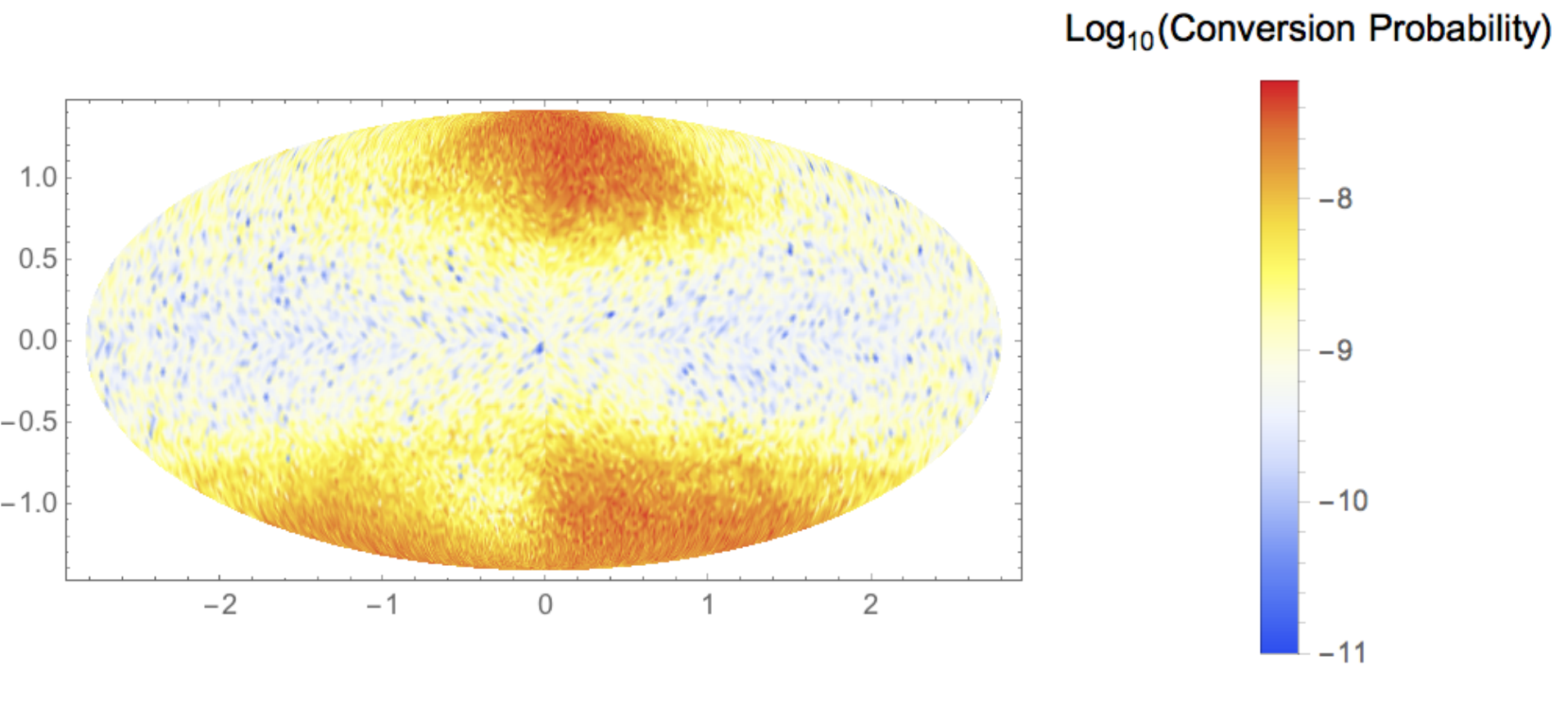}
\caption{The ALP to photon conversion probability in the Milky Way for ALP energy $\omega = 200 \, {\rm eV}$ and $M = 10^{13} \, {\rm GeV}$.}
\label{200Prob}
\end{figure}

\begin{figure} [h]
\includegraphics[scale=0.7]{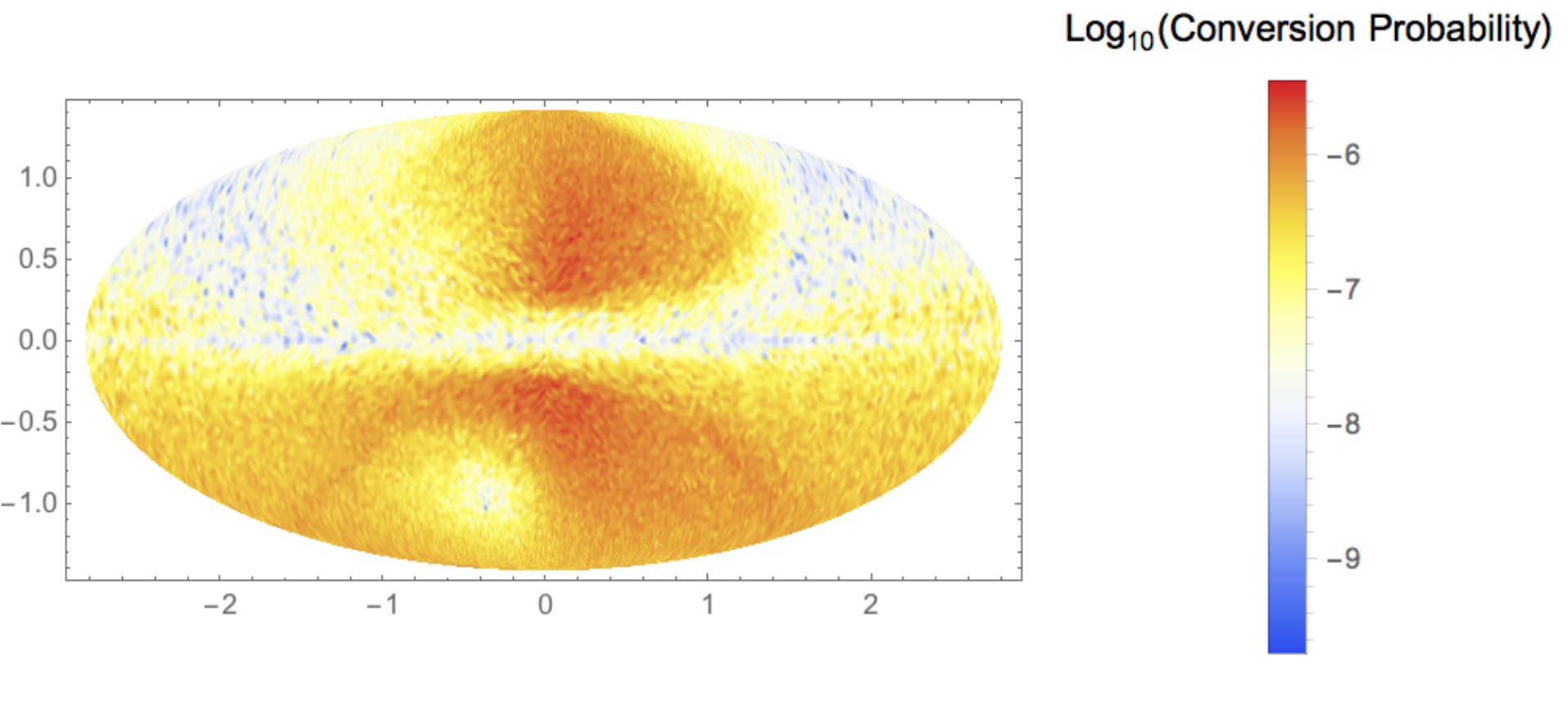}
\caption{The ALP to photon conversion probability in the Milky Way for ALP energy $\omega = 500 \, {\rm eV}$ and $M = 10^{13} \, {\rm GeV}$.}
\label{500Prob}
\end{figure}

\begin{figure} [h]
\includegraphics[scale=0.7]{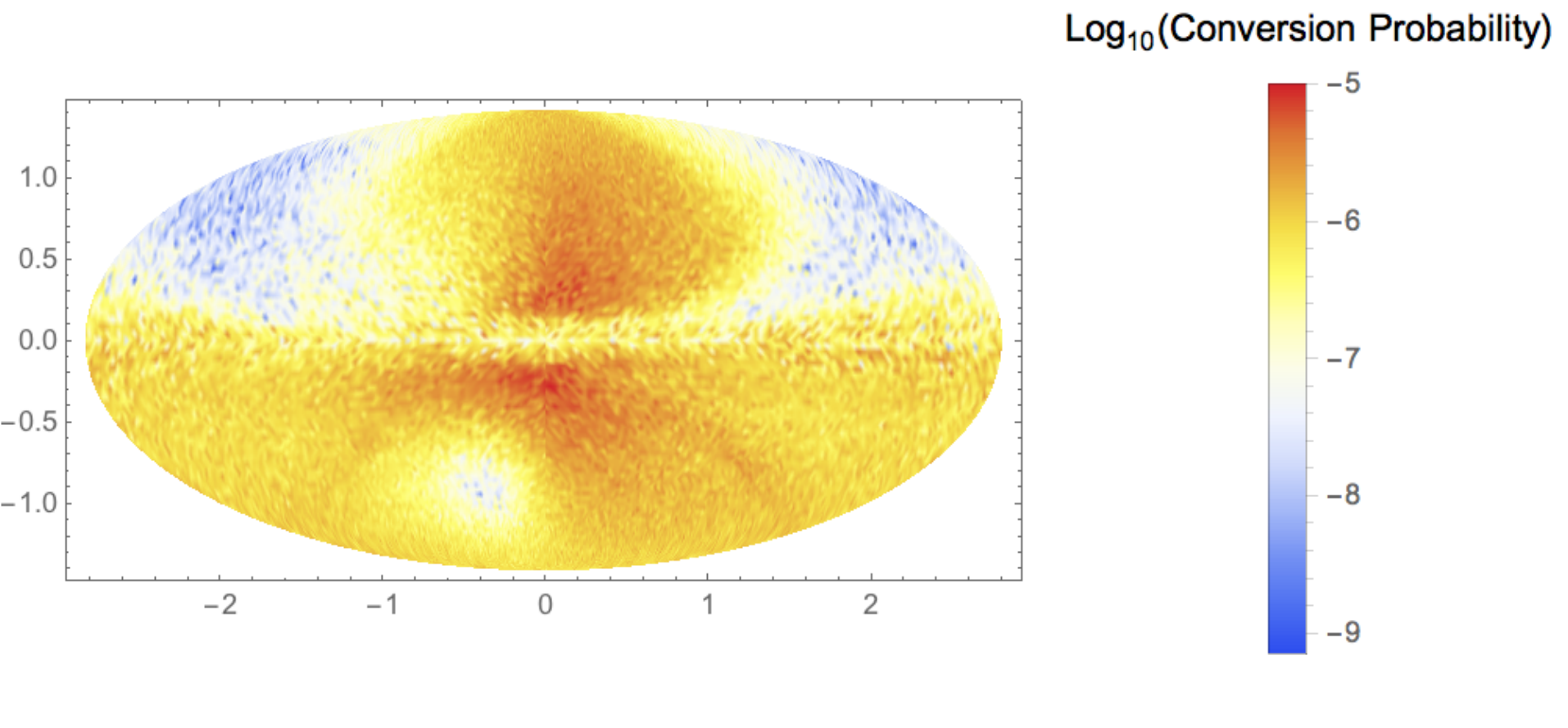}
\caption{The ALP to photon conversion probability in the Milky Way for ALP energy $\omega = 1500 \, {\rm eV}$ and $M = 10^{13} \, {\rm GeV}$.}
\label{1500Prob}
\end{figure}

The ALP to photon conversion probability across the Milky Way is relevant to a range of effects, probing different ALP energies. Our focus in this paper is on the propagation of a cosmic ALP background at soft X-ray energies converting to photons in the Milky Way.  The Milky Way ALP to photon conversion probability at a range of soft X-ray energies is shown in figures \ref{200Prob}, \ref{500Prob} and \ref{1500Prob}. We see that at $\omega = 200 \, {\rm eV}$, $P_{a \to \gamma}$ is heavily suppressed by photoelectric absorption but this suppression is not significant at  $\omega = 500 \, {\rm eV}$, where $P_{a \to \gamma}$ inherits the geometry of the galactic magnetic field.

\begin{figure} [h]
\includegraphics[scale=0.7]{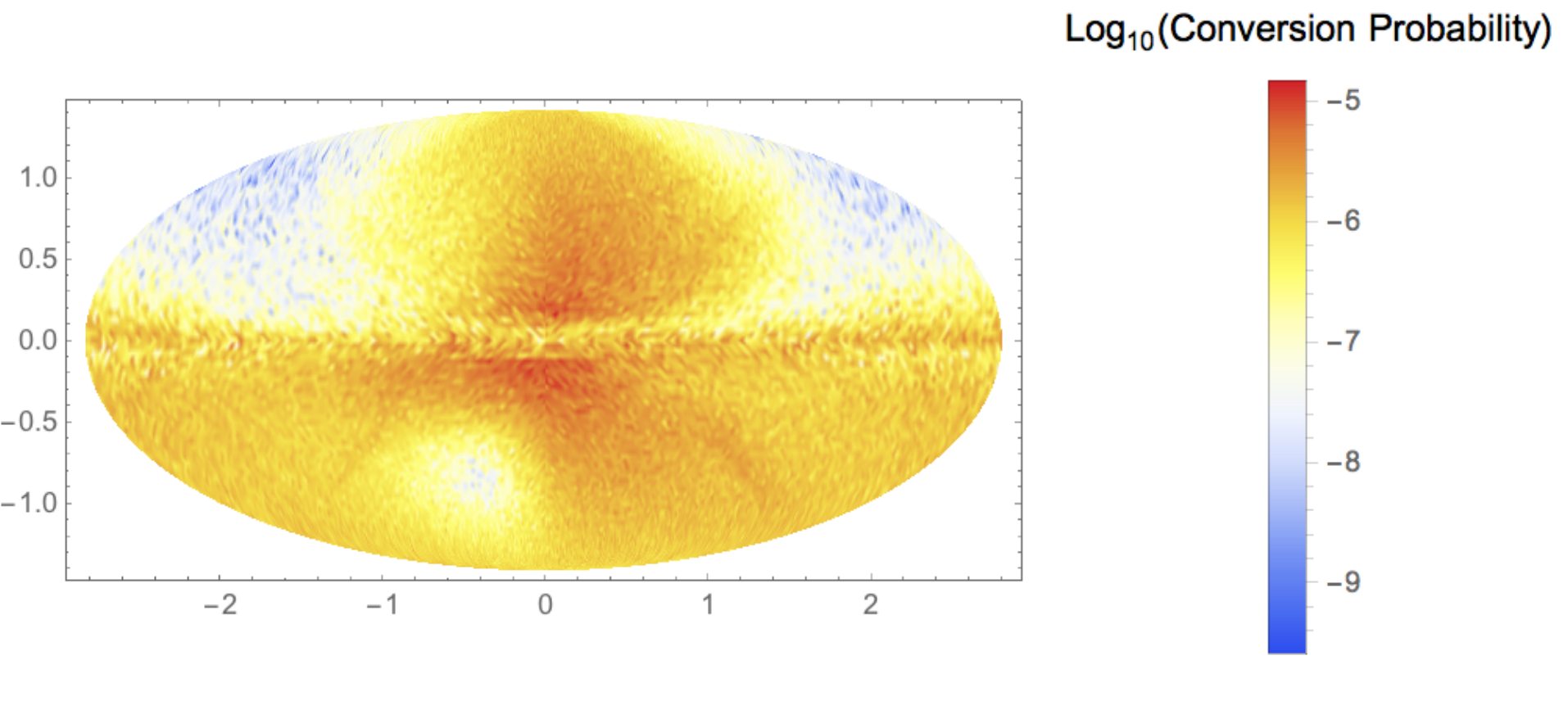}
\caption{The ALP to photon conversion probability in the Milky Way for ALP energy $\omega = 3.5 \, {\rm keV}$ and $M = 10^{13} \, {\rm GeV}$.}
\label{3500Prob}
\end{figure}

It has been suggested that the 3.5 keV line recently observed in galaxy clusters and the Andromeda galaxy may arise from dark matter decay to ALPs followed by ALP-photon conversion in astrophysical magnetic fields \cite{14032370,14047741,14065518,14101867}. This scenario fits the morphology of the 3.5 keV line flux in galaxy clusters and predicted its non-observation in a stacked sample of galaxies. In this case, the conversion probability for 3.5 keV ALPs in the Milky Way, shown in figure \ref{3500Prob} is required. As shown in \cite{14047741}, the conversion probability is too low to expect an observable 3.5 keV line flux in the Milky Way halo.

\begin{figure} [h]
\includegraphics[scale=0.7]{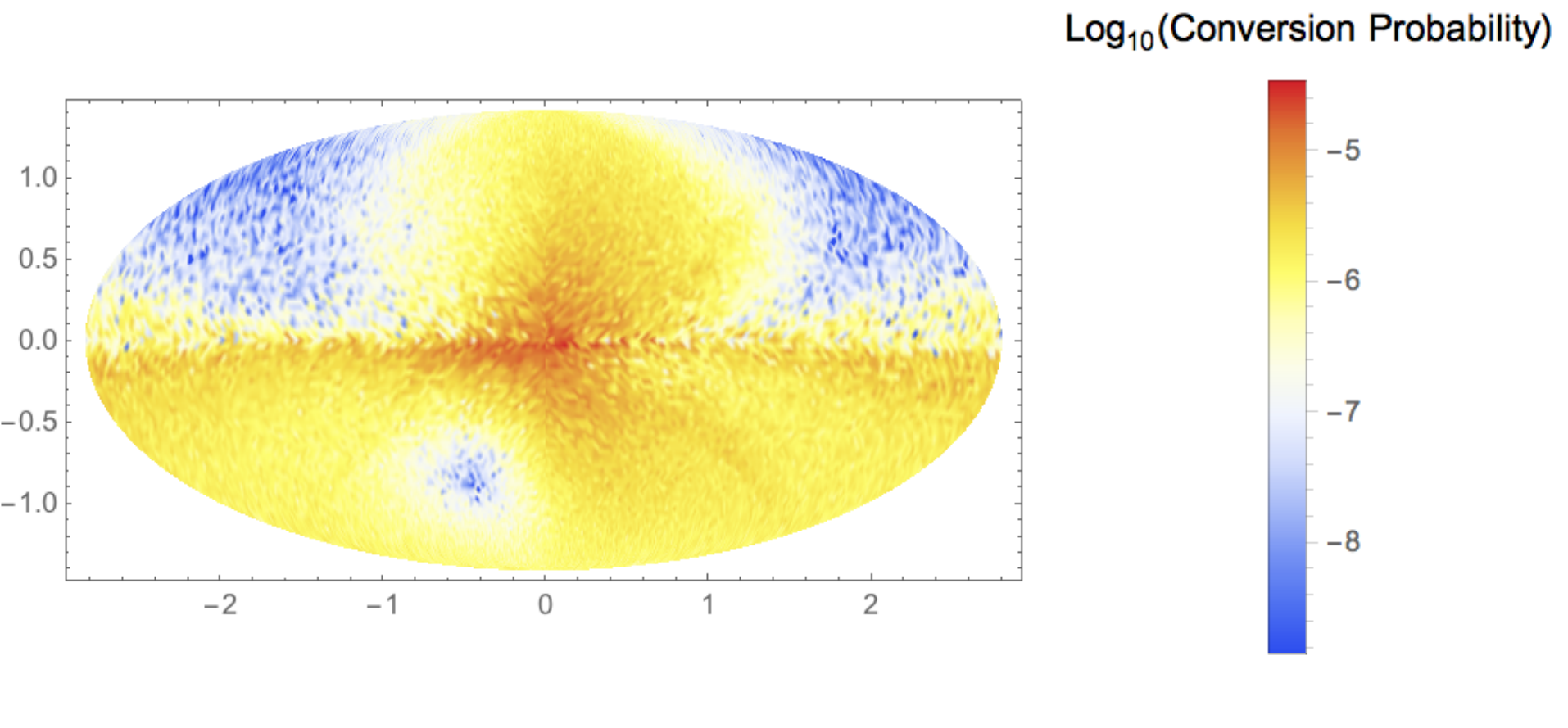}
\caption{The ALP to photon conversion probability in the Milky Way for ALP energy $\omega = 1 \, {\rm TeV}$ and $M = 10^{13} \, {\rm GeV}$.}
\label{TeVProb}
\end{figure}

ALP-photon conversion has also been suggested as an explanation for the anomalous transparency of the universe to gamma rays \cite{07043044,07074312,Simet,09053270,12070776,13021208,Wouters,14060239}. In this scenario, gamma rays emitted by distant blazars convert to ALPs in the magnetic field of the host galaxy or in the intergalactic magnetic field, and then reconvert to photons in the intergalactic or Milky Way magnetic field. In this way, gamma ray photons are able to avoid scattering from electrons in intergalactic space. In this case, the conversion probability for gamma ray energy ALPs, shown in figure \ref{TeVProb}, is key. This conversion probability is also used in calculating the bounds on $M$ from SN1987a. Comparing figures \ref{1500Prob}, \ref{3500Prob} and \ref{TeVProb}, we see that the conversion probability in the Milky Way saturates at $P_{a \to \gamma} \sim 10^{-6}$ for $M = 10^{13} \, {\rm GeV}$. This behaviour can be seen in the single domain formula (equation \eqref{singleDomain}), where for $\omega \to \infty$, the analytic formula for reduces to:

\be 
\lim_{\omega \to \infty} P_{a \to \gamma} = {\rm sin}^{2}\left(\frac{L B_{\perp}}{2 M}\right).
\ee

\subsection{Application to a cosmic ALP background}
\label{CAB}

A cosmic ALP background would convert to photons in astrophysical magnetic fields leading to a potentially observable soft X-ray flux. This effect has been suggested as the source of the soft X-ray excess in galaxy clusters. The ALP to photon conversion probability in the Milky Way is around 3 orders of magnitude lower than that in galaxy clusters, primarily due to the Milky Way's smaller size. We therefore do not expect such a strong signal from CAB to photon conversion in the Milky Way. Any extra soft X-ray photons generated from a CAB's passage through the Milky Way would contribute to the unresolved cosmic X-ray background - the diffuse soft X-ray flux remaining after subtracting the flux from known point sources.

\begin{figure} [H]
\includegraphics[scale=0.7]{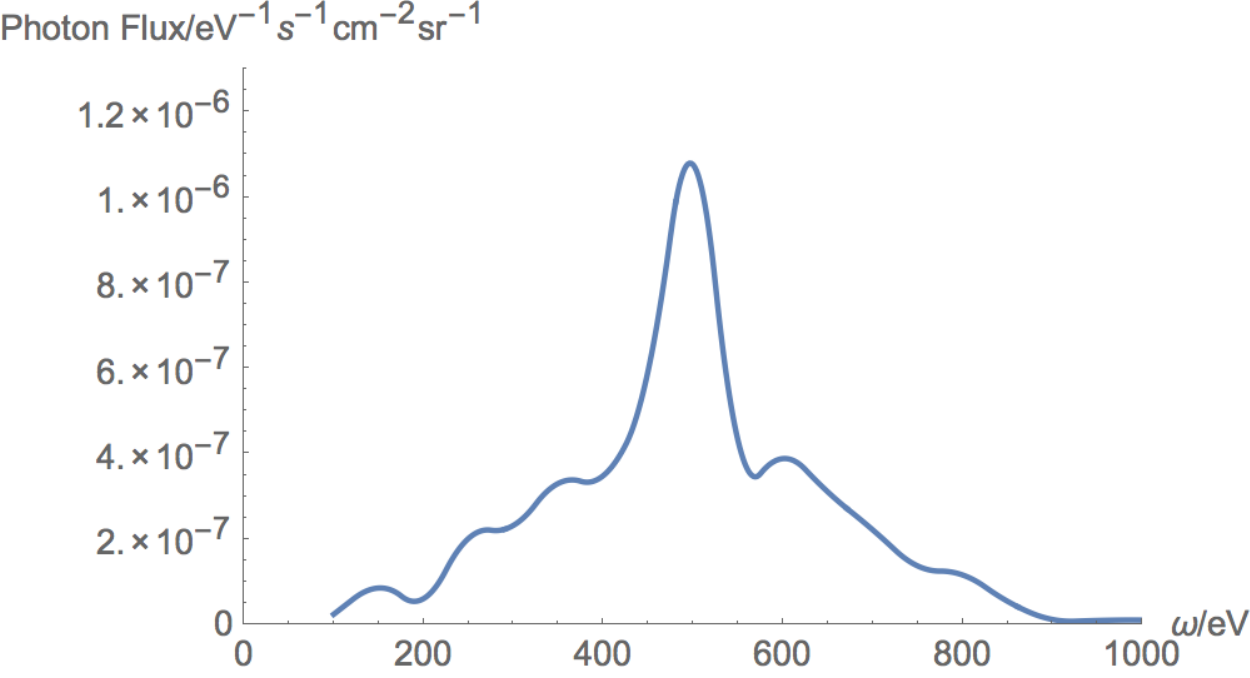}
\caption{The predicted photon flux from the Milky Way centre from CAB to photon conversion with $E_{\rm CAB} = 200 \, {\rm eV}$ and $M = 10^{13} \, {\rm GeV}$}
\label{photonFluxDisc}
\end{figure}

\begin{figure} [H]
\includegraphics[scale=0.7]{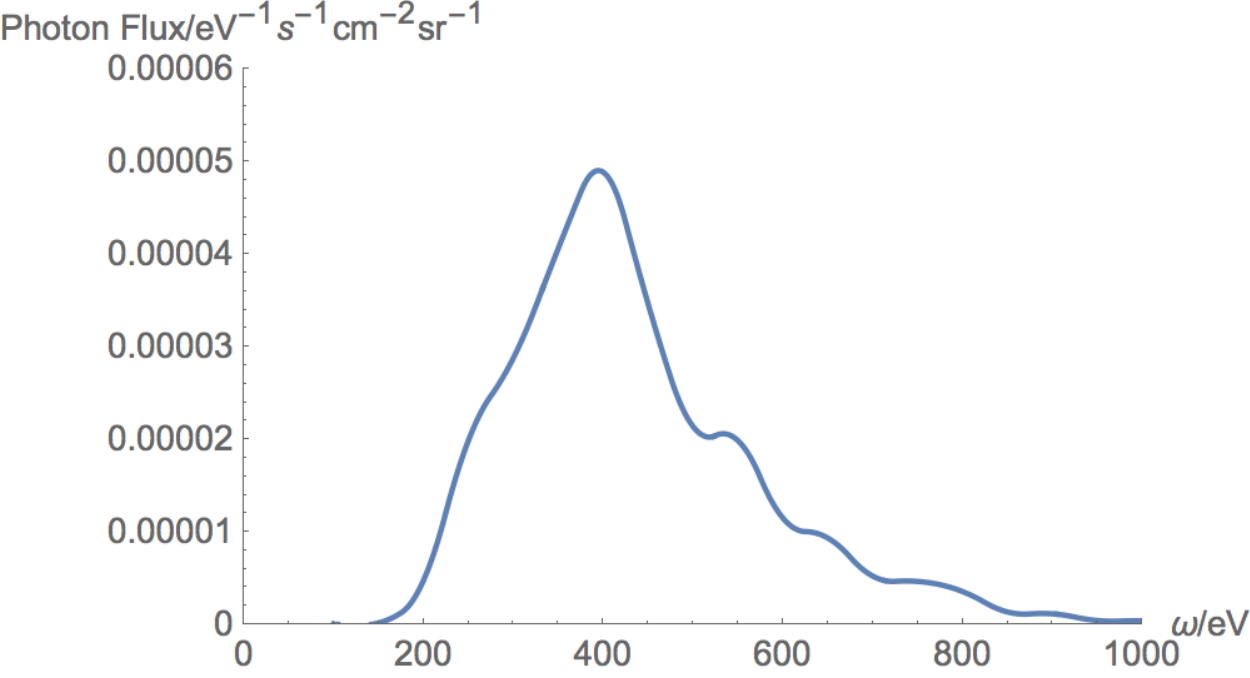}
\caption{The predicted photon flux from due Galactic North from CAB to photon conversion with $E_{\rm CAB} = 200 \, {\rm eV}$ and $M = 10^{13} \, {\rm GeV}$}
\label{photonFluxHalo}
\end{figure}

The photon flux from a CAB in a solid angle $\Omega$ is given by:

\be 
\label{flux}
\frac{d \Phi_{\Omega}}{dE} = \frac{1}{2 \pi}\int_{\Omega} d \Omega^{\prime} P\left( \Omega^{\prime}, E \right) A X\left( E \right) \frac{c}{4}.
\ee
For example, for a central energy $E_{\rm{CAB}} = 200 \, \rm{eV}$, the predicted photon fluxes for disc and halo pointings are shown in figures \ref{photonFluxDisc} and \ref{photonFluxHalo}. We normalise to $\Delta N_{{\rm eff}} = 0.5$ to allow easy comparison with the galaxy cluster fluxes simulated in \cite{13123947}. The predicted CAB signal scales linearly with the CAB contribution to $\Delta N_{{\rm eff}}$. We notice that the shape of the spectrum is dramatically altered from the CAB spectrum shown in figure \ref{ALPflux}, as the conversion probability at low energies is dramatically suppressed by photo-electric absorption. The spectrum is further influenced by the energy dependence of the conversion probability even in the absence of absorption. For example, for both pointings we see oscillations in the predicted flux on top of the overall shape of the spectrum. These can be understood by considering the analytic solution in equation \eqref{singleDomain}, which approximates the qualitative features of the solution in the non-homogeneous case simulated here. In particular, we expect to see local maxima in the conversion probability whenever $\Delta = 0.053 \times \left( \frac{n_{e}}{10^{-3} \, {\rm cm}^{-3}} \right) \left(\frac{1 \, {\rm keV}}{\omega} \right) \left( \frac{L}{1 \, {\rm kpc}} \right) = \frac{N \pi}{2}$ for odd integer $N$. These correspond to the oscillations seen in figures \ref{photonFluxDisc} and \ref{photonFluxHalo} and are a distinctive feature of a photon flux arising from ALP to photon conversion in a sufficiently high electron density environment (so that $\Delta > 1$). The flux from the Milky Way centre (figure \ref{photonFluxDisc}) is lower and peaks at higher energies that that from due Galactic North (figure \ref{photonFluxHalo}) due to the higher warm neutral medium column density towards the Galactic centre, leading to a greater effect from photo-electric absorption. Note that the detailed shape of the Milky Way centre spectrum is highly dependent on the realisation of the strong random magnetic field in the disc.

\begin{figure} [H]
\includegraphics[scale=0.7]{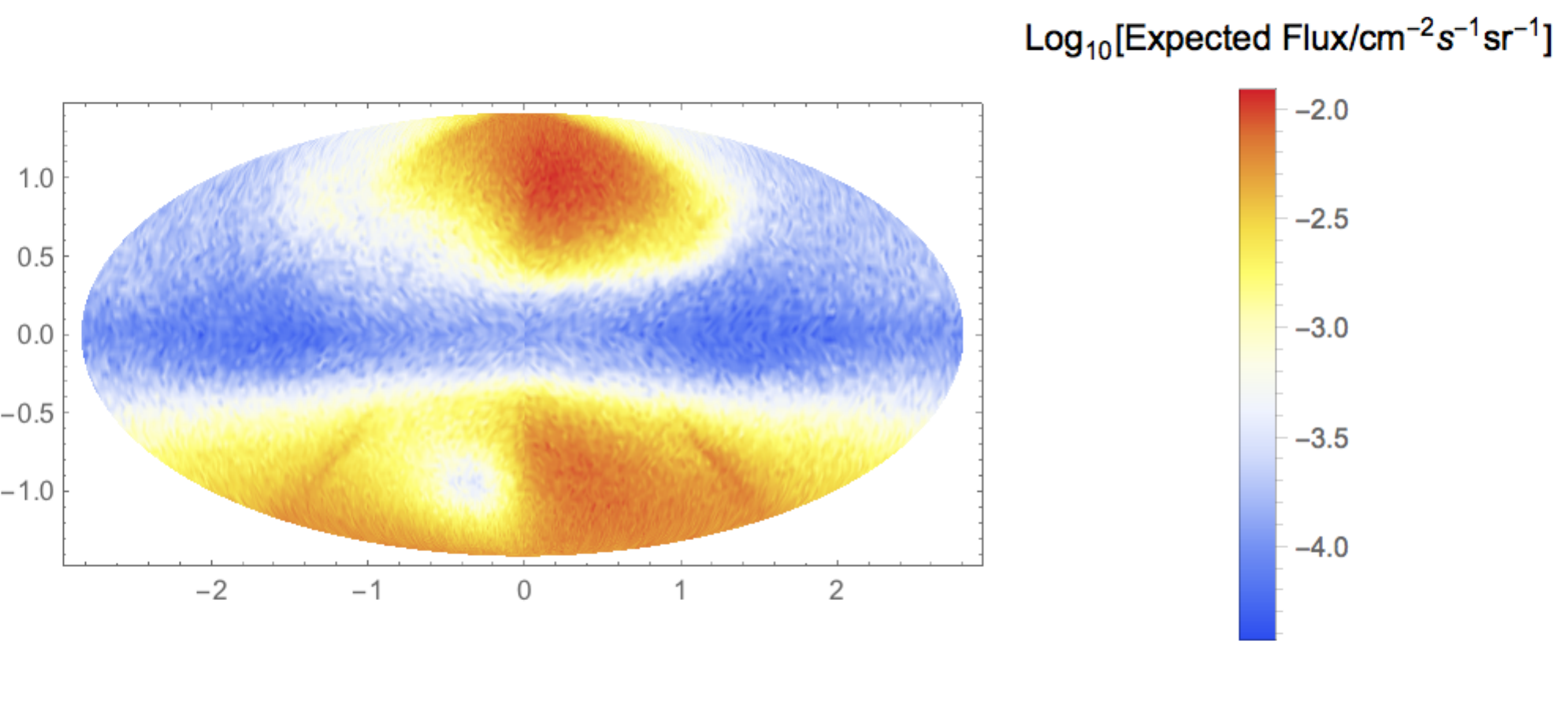}
\caption{The predicted photon flux in the ROSAT 1/4 keV band for $E_{{\rm CAB}} = 200 \, {\rm eV}$, $M = 5 \times 10^{12} \, {\rm GeV}$ and $\Delta N_{{\rm eff}} = 0.5$}
\label{quarterBand}
\end{figure}

\begin{figure} [H]
\includegraphics[scale=0.7]{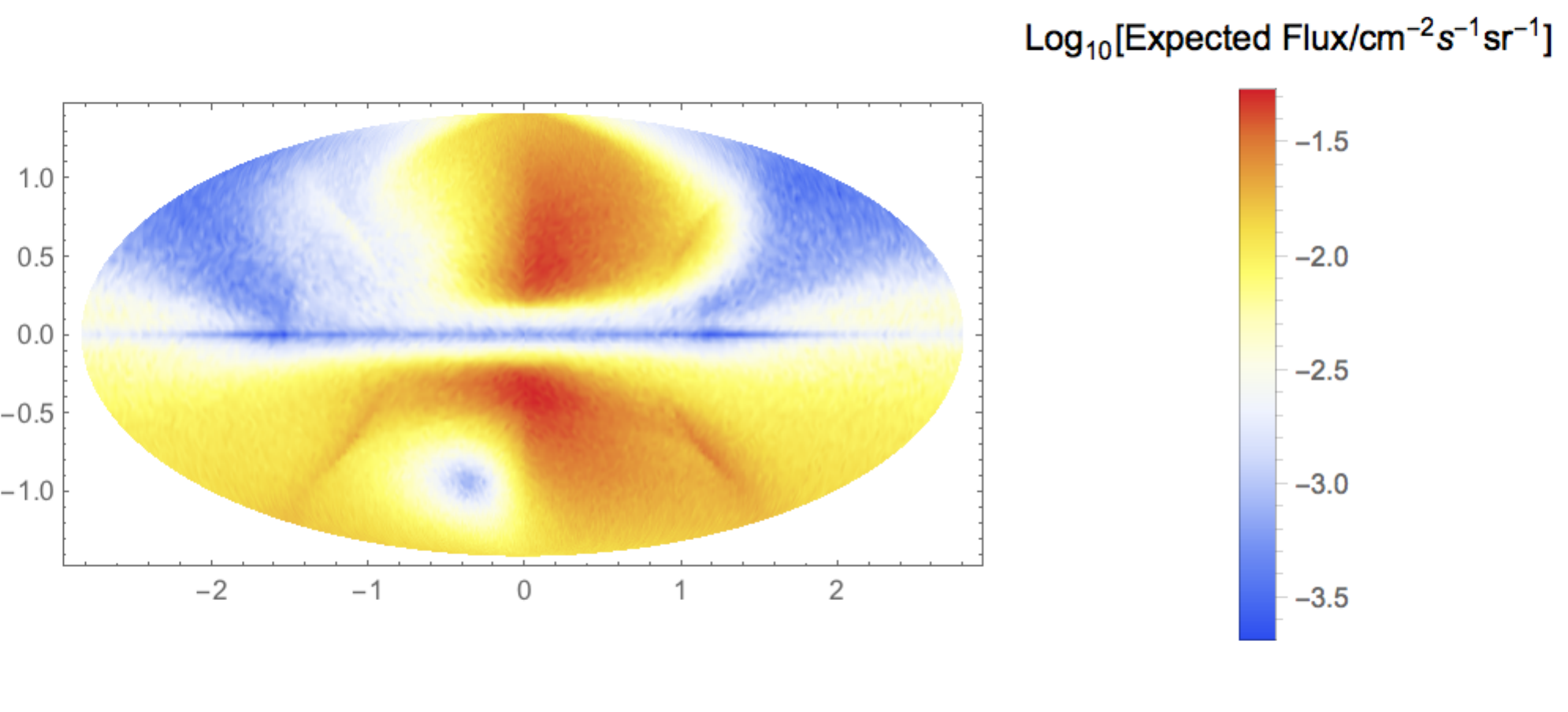}
\caption{The predicted photon flux in the ROSAT 3/4 keV band for $E_{{\rm CAB}} = 200 \, {\rm eV}$, $M = 5 \times 10^{12} \, {\rm GeV}$ and $\Delta N_{{\rm eff}} = 0.5$}
\label{threequarterBand}
\end{figure}

Full sky maps of the cosmic X-ray background were observed by the ROSAT satellite \cite{ROSAT}. We now calculate the predicted flux from a CAB in the ROSAT 1/4 keV and 3/4 keV bands. We use equation \eqref{flux} with the conversion probabilities calculated using equation \eqref{discrete}. As shown in \cite{13123947,outskirts,14114172}, natural CAB parameters to explain the cluster soft excess are $E_{{\rm CAB}} = 200 \, {\rm eV}$ and $M = 5 \times 10^{12} \, {\rm GeV}$ for $\Delta N_{{\rm eff}} = 0.5$. We plot the predicted ROSAT signals for these parameters as full sky maps in figures \ref{quarterBand} and \ref{threequarterBand}. Comparing with \cite{ROSAT}, we find that the predicted CAB signal is over 3 orders of magnitude smaller than the signal observed by ROSAT. The soft X-ray excess can also be explained with a lower $E_{{\rm CAB}}$ and lower $M$ - in this case the signal in the Milky Way is even lower due to the higher photo-electric absorption at lower energies.

We might wonder if a CAB with different parameters could contribute significantly to the cosmic X-ray background (quite possibly by ignoring the problem of overproduction in clusters). The cosmic X-ray background is most clearly seen in the Chandra Deep Field (CDF) observations \cite{CDF}. The observed fluxes, ALP to photon conversion probabilities and predicted fluxes from CAB to photon conversion (with the parameters used above) for the CDF-South and CDF-North observations are shown in table 1. To simulate the conversion probabilities here, we did not include the effects of photo-electric absorption, as the CDF pointings are chosen for their low warm neutral medium column density. For any CAB parameter values we expect the CAB signal from CDF-North to be $\mathcal{O} \left( 10 \right)$ times lower than that from CDF-South. However, the cosmic X-ray background intensities from these observations are the same within their errors. Therefore the possibility of a CAB forming the dominant part of the cosmic X-ray background is excluded by the North-South asymmetry of the Milky Way magnetic field.

\begin{table}[h]
\begin{tabular}{|l|l|l|l|l|}
\hline
& $(b,l)$ &  $P_{a \to \gamma}$  & Predicted Flux & Observed Flux \\
\hline
CDF-South &  $(-22.8^{\circ}, 161^{\circ})$    & $ 2.4 \times 10^{-6}$ & $2.4 \times 10^{-15}$ & $\left(1.1 \pm 0.2 \right) \times 10^{-12}$ \\
\hline
CDF-North & $(54.8^{\circ}, 125.9^{\circ})$    & $2.5 \times 10^{-7}$ & $2.0 \times 10^{-16}$ & $\left(9 \pm 3 \right)\times 10^{-13}$ \\
\hline
\end{tabular}
\caption{The ALP to photon conversion probabilities $P_{a \to \gamma}$ (averaged over the energy range), predicted CAB fluxes and observed cosmic X-ray background fluxes after point source subtraction for the Chandra Deep Field pointings. Fluxes are given in units of ${\rm ergs} \, {\rm cm}^{-2} \, {\rm s}^{-1}\, {\rm deg}^{-2}$. We also show the galactic latitude $b$ and longitude $l$ of the observations. We use $E_{{\rm CAB}} = 200 \, {\rm eV}$, $M = 5 \times 10^{12} \, {\rm GeV}$ and $\Delta N_{{\rm eff}} = 0.5$. }
\end{table}


We see that a CAB responsible for the cluster soft X-ray excess would currently be unobservable in the Milky Way, and that a CAB cannot contribute significantly to the observed unresolved cosmic X-ray background without giving it a North-South asymmetry ruled out by observations. Ubiquitous features of a CAB Milky Way signal are a prominent North-South asymmetry (as shown in figures \ref{quarterBand} and \ref{threequarterBand}), and complex spectral shapes from a convolution of the quasi-thermal CAB spectrum and the energy dependent conversion probability as shown in figures \ref{photonFluxDisc} and \ref{photonFluxHalo}. In particular, the conversion probability and therefore the predicted flux oscillates as the energy increases.

\section{Additional effects}

\subsection{Milky Way magnetic field}

We have used a simplistic model for the random and striated fields with a single coherence length of 100 pc. In reality, we expect these fields to exhibit a range of coherence scales. However, changing the coherence length by a factor of 10 in either direction only results in a factor of $\lesssim 2$ difference in the full conversion probability. Furthermore, we have not considered the field at the very centre of the Milky Way. ALP to photon conversion in the Milky Way centre is discussed in \cite{14101867} in the context of the 3.5 keV line. Here we simply note that estimates of the magnetic field in the Galactic centre are highly uncertain, ranging from $10 \, \mu {\rm G}$ to $1 \,  {\rm mG}$. At the upper end of this field range, we could see conversion probabilities in the Milky Way centre comparable to those in galaxy clusters, and therefore might expect an observable soft X-ray flux from a CAB. However, the high density of the warm neutral medium in the Galactic centre would significantly suppress the signal at low energies. Furthermore, the galactic centre is a highly complex environment so it would be very difficult to pick out a small excess soft X-ray flux.  

\subsection{Clumpiness of the interstellar medium}

\begin{figure} [H]
\begin{center}
\includegraphics[width=0.45\textwidth]{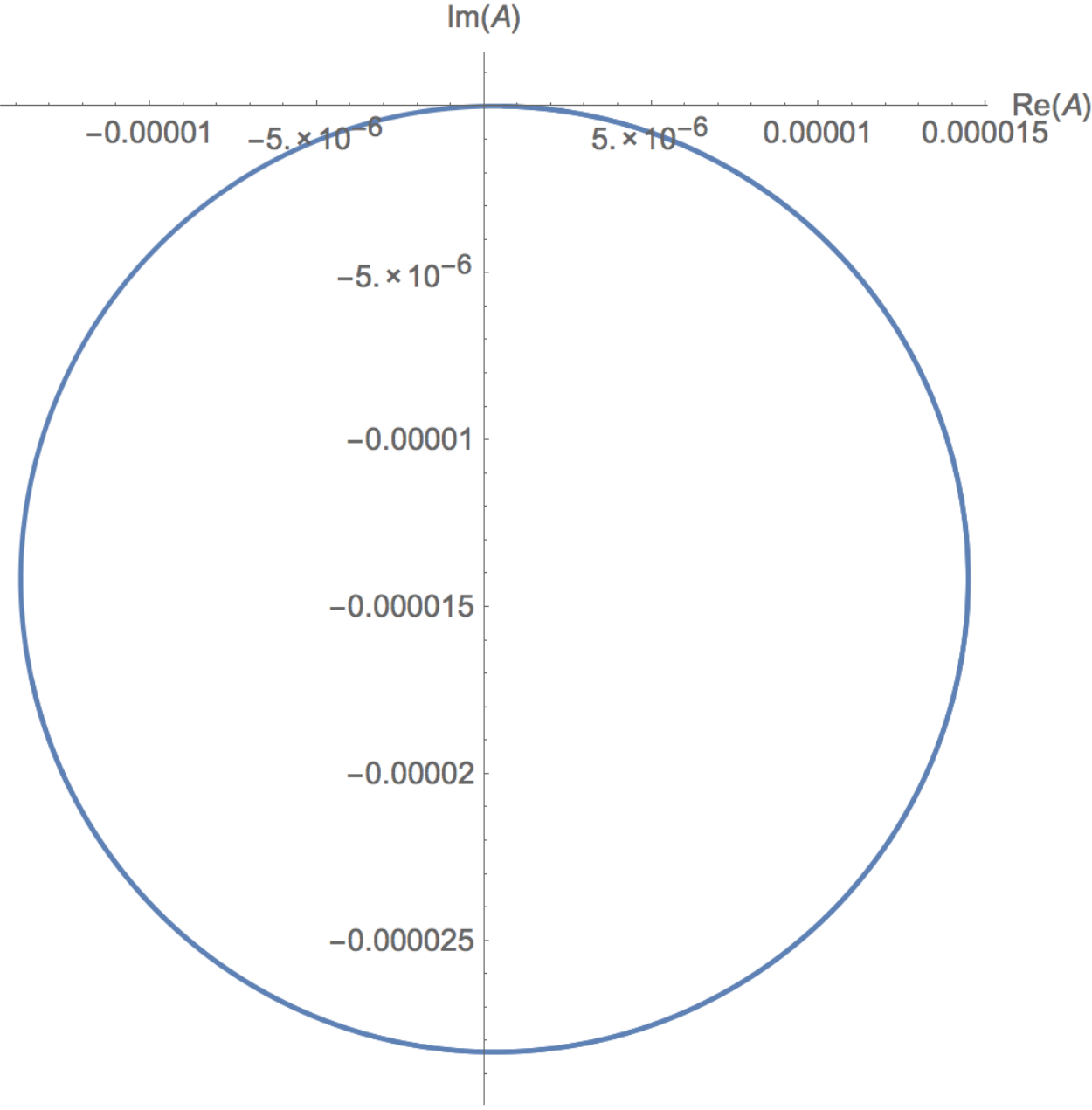}~~~\includegraphics[width=0.45\textwidth]{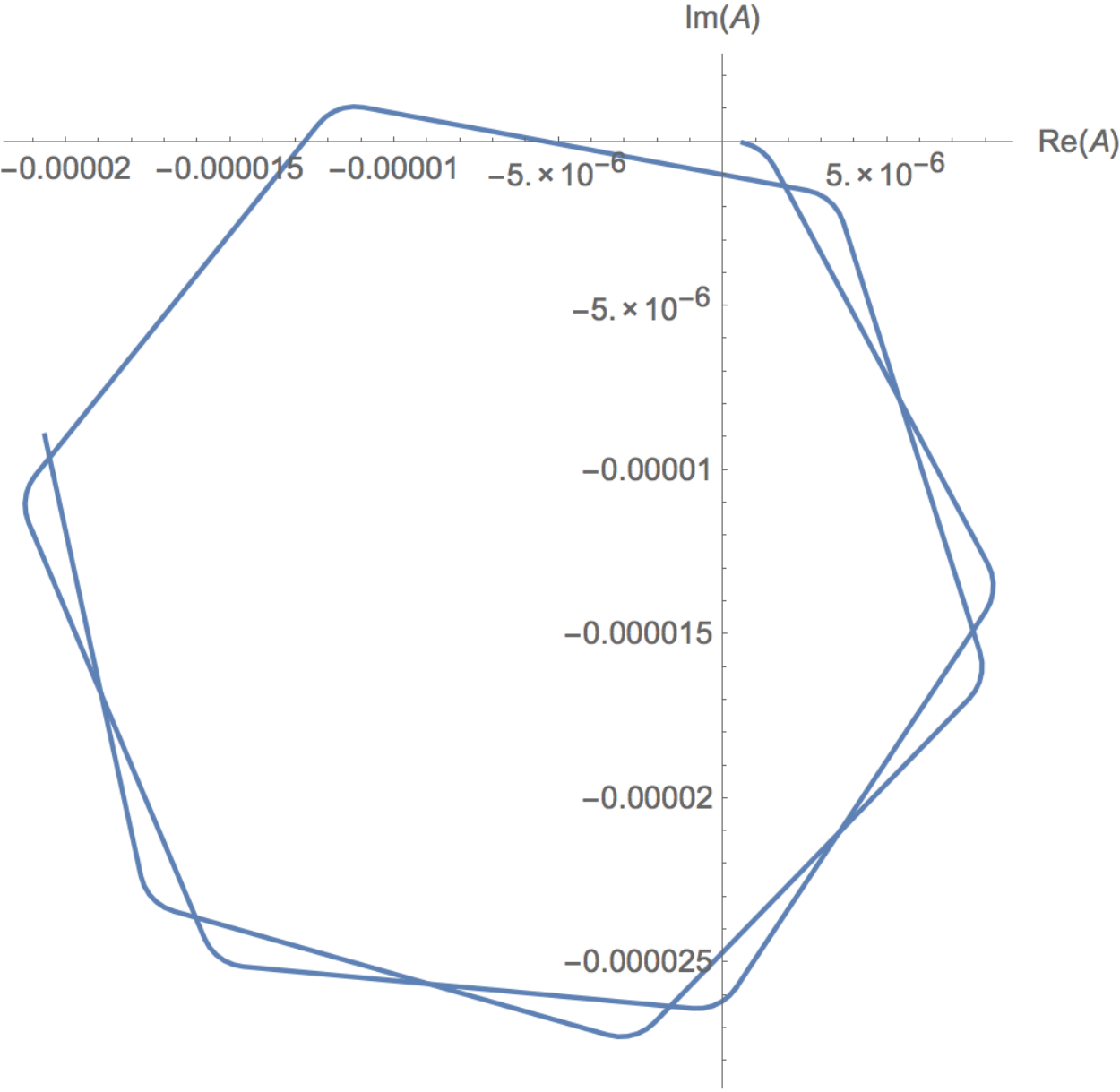}
\end{center}
\caption{$A(L) = \int^{L}_{0} dz e^{i \varphi(z )} \Delta_{\gamma a i}(z)$ for $L = 0 - 1 \, {\rm kpc}$ increasing along the line. In the left hand plot we use the volume averaged electron density, and in the right hand plot implement the electron density in evenly spaced clouds, with an intercloud electron density of $10^{-7} \, {\rm cm}^{-3}$. The clouds correspond to the `corners' in the plot. We see that $P_{a \to \gamma} = |A|^2$ is not significantly effected by the presence of clouds. We use $B_{\perp} = 1 \, \mu {\rm G}$, volume average electron density $\overline{n}_{e} = 0.05 {\rm cm}^{-3}$, cloud size $d_{c} = 10 \, {\rm pc}$, filling factor $f = 0.1$, $M = 10^{13} \, {\rm GeV}$ and $\omega = 500 \, {\rm eV}$.  For the volume averaged case, $P_{a \to \gamma} \left( L = 1 \, {\rm kpc} \right) = 4.9 \times 10^{-10}$. With clouds, $P_{a \to \gamma} \left( L = 1 \, {\rm kpc} \right) = 5.2 \times 10^{-10}$.}
\label{low}
\end{figure}

We recall that the electron density of the surrounding medium suppresses ALP-photon conversion by giving an effective mass to the photon, as shown in equations \eqref{propfree}, \eqref{Deltagamma}, \eqref{semi-analytic} and \eqref{phi}. The electron density model used to simulate ALP-photon conversion describes the smooth, \emph{volume averaged} electron density. In reality, the warm ionized medium (WIM) in galaxies exists in high density clouds with a rather low intercloud electron density \cite{Berkhuijsen}. This structure is characterized by the filling factor $f$, the fraction of a line of sight occupied by WIM clouds. In principle, by using the clumpy local electron density $n_{e}$ we might predict a different $P_{a \to \gamma}$ than we would have by naively implementing the smooth volume averaged electron density $\overline{n}_{e}$. To examine the effect of the local electron density distribution, we consider the role of the electron density in rotating the probability amplitude $A(L) = \braket{1,0,0|f(L)}$ in the complex plane. For simplicity, we consider the case of a constant magnetic field in the $x$ direction, so that only $x$ polarized photons are produced. The relevant equations are then (see equations \eqref {semi-analytic} and \eqref{phi}):

\be
P_{a\to \gamma}(L)=\left|\int^{L}_{0} dz e^{i \varphi(z )} \frac{B_{x}(z)}{2M}\right|^2 \, ,
\ee
where,
\be
\varphi(z)  = \int^{z}_{0} dz' \Delta_{\gamma}(z') = - \frac{1}{2\omega} \int^{z}_{0} dz' \omega_{pl}^2(z') \, ,
\ee
with 
\be 
\omega_{pl}^{2} =  4 \pi \alpha \frac{n_{e}}{m_{e}}.
\ee
We see that the angle of turn in the complex plane is given by $\varphi(z )$, which is linear in $n_{e} \left( z \right)$. We first note that whether this turning happens continuously or in steps does not significantly effect $P_{a \to \gamma}$. This is demonstrated in figure \ref{low}, where we plot in the complex plane the probability amplitude $A(L) = \int^{L}_{0} dz e^{i \varphi(z )} \Delta_{\gamma a i}(z)$ for a propagation distance $L = 0 - 1 \, {\rm kpc}$ increasing along the line. We use $B_{\perp} = 1 \, \mu {\rm G}$, $\overline{n}_{e} = 0.05 \, {\rm cm}^{-3}$, cloud size $d_{c} = 10 \, {\rm pc}$, $f = 0.1$ and $\omega = 500 \, {\rm eV}$. In the left hand plot we use the volume averaged electron density, and in the right hand plot implement the electron density in evenly spaced clouds, with an intercloud electron density of $10^{-7} \, {\rm cm}^{-3}$. We see that the overall shape of $A(L)$ is the same in each case, although in the volume averaged case the turn is continuous, whereas in the right hand plot we see corners (with a very high rate of turn) where there is a cloud. As expected, the conversion probability at $L = 1 \, {\rm kpc}$ is practically the same in each case ($4.9 \times 10^{-10}$ and $5.2 \times 10^{-10}$ respectively for $M = 10^{13} \, {\rm GeV}$).

\begin{figure} [H]
\begin{center}
\includegraphics[width=0.45\textwidth]{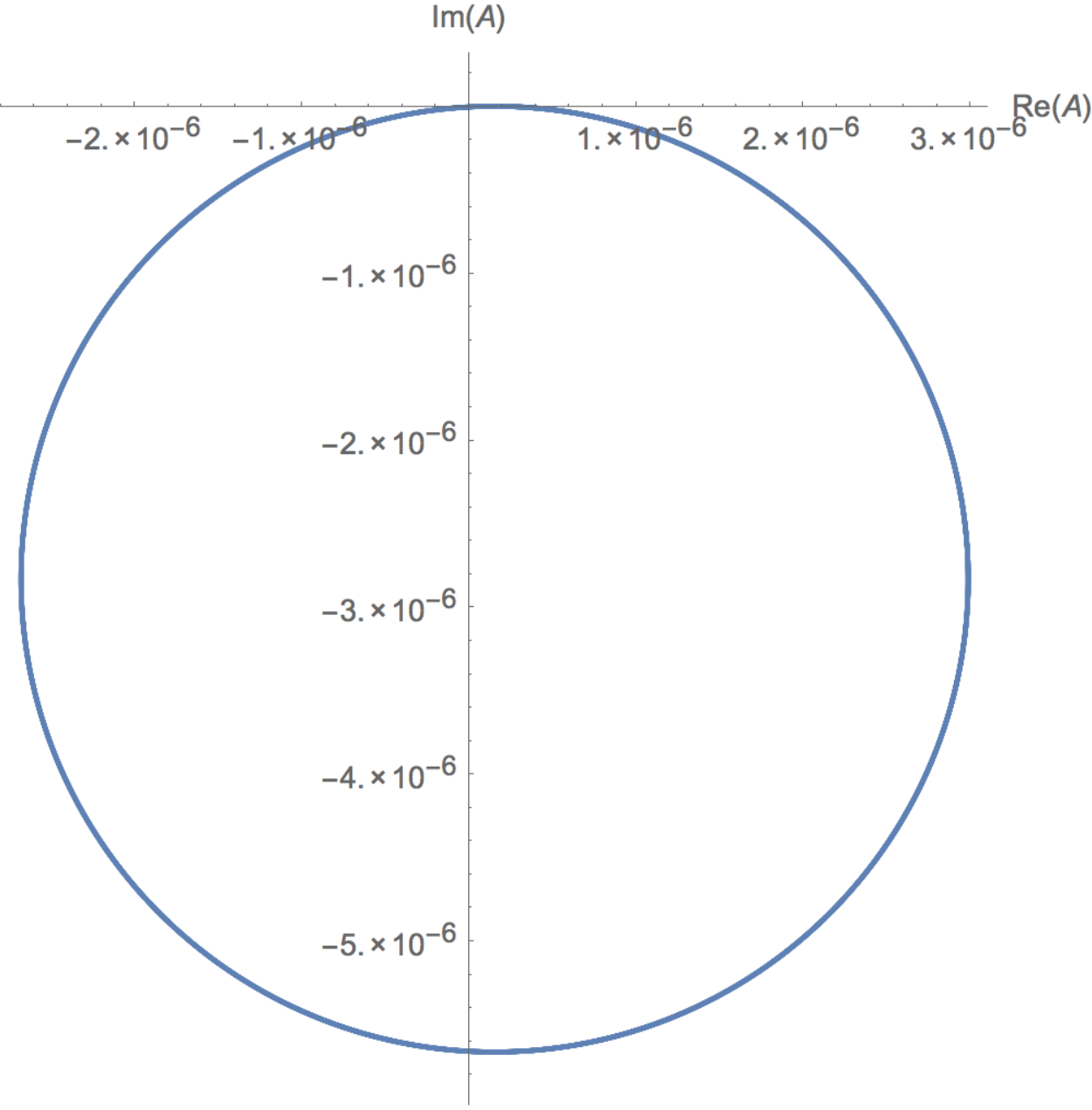}~~~\includegraphics[width=0.45\textwidth]{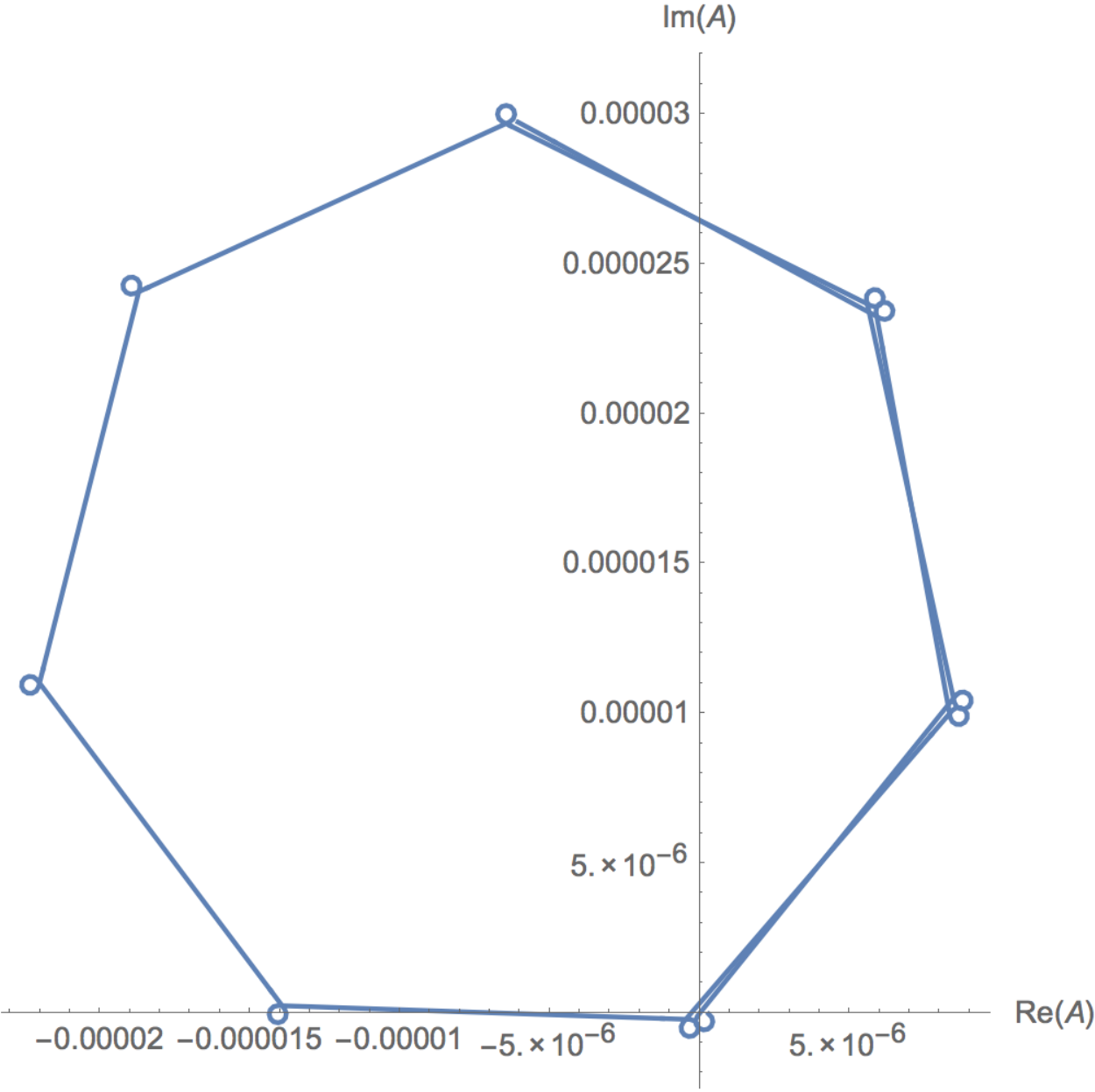}
\end{center}
\caption{$A(L) = \int^{L}_{0} dz e^{i \varphi(z )} \Delta_{\gamma a i}(z)$ for $L = 0 - 1 \, {\rm kpc}$ increasing along the line.  In the left hand plot we use the volume averaged electron density, and in the right hand plot implement the electron density in evenly spaced clouds, with an intercloud electron density of $10^{-7} \, {\rm cm}^{-3}$. The clouds correspond to the loops in the plot. We see that $P_{a \to \gamma} = |A|^2$ is significantly higher when the cloud structure is taken into account. We use $B_{\perp} = 1 \, \mu {\rm G}$, volume average electron density $\overline{n}_{e} = 0.05 {\rm cm}^{-3}$, cloud size $d_{c} = 10 \, {\rm pc}$, filling factor $f = 0.1$, $M = 10^{13} \, {\rm GeV}$ and $\omega = 100 \, {\rm eV}$. For the volume averaged case, $P_{a \to \gamma} \left( L = 1 \, {\rm kpc} \right) = 3.0 \times 10^{-11}$. With clouds, $P_{a \to \gamma} \left( L = 1 \, {\rm kpc} \right) = 9.2 \times 10^{-10}$.}
\label{high}
\end{figure}

However, $e^{i \varphi(z )}$ is also \emph{periodic} in $n_{e} \left( z \right)$, and it is in this periodicity that we see the effect of the clumpiness of the WIM. For high electron densities, low filling factors and/or low ALP energies it may be that within a single cloud $ \varphi(z )$ changes by $\gtrsim 2 \pi$.  A cycle of ${\rm Arg} \left( e^{i \varphi(z )} \right) $ within a cloud that returns to its starting point does not significantly decrease the overall conversion probability. In the regime where this phenomenon occurs, the overall \emph{large scale} turning of $A$ in the complex plane is decreased by the organisation of the WIM into clouds. The significant ($\gtrsim 2\pi$) turning within a cloud essentially gives us a `free lunch' - the volume averaged electron density is increased, but there is no contribution to the net large scale turning, and so the overall conversion probability is not significantly decreased. The predicted conversion probability is therefore significantly increased by taking into account the cloud structure of the WIM. This effect is demonstrated by figure \ref{high}. Here we use the same parameters as in figure \ref{low}, but with $\omega = 100 \, {\rm eV}$ (so that $\Delta_{\gamma}$ is increased by a factor of 5). We now see the turns within clouds in the right hand plot, allowing $A(L)$ to reach larger radii in the complex plane, and increasing $P_{a \to \gamma}(L)$ from $3.0 \times 10^{-11}$ with a smooth volume averaged WIM to $9.2 \times 10^{-10}$ with a more realistic WIM profile. In spite of the very high electron density within clouds $n_{c} = \frac{\overline{n}_{e}}{f}$, the low electron density intercloud regions allow $|A(L)|$ to grow in this regime.

We therefore see that the condition for the cloud structure of the WIM to be significant is:

\be
\label{delta}
\delta = 1.1 \times 10^{-2} \left( \frac{n_{c}}{10^{-2} \, {\rm cm}^{-3}} \right) \left( \frac{{\rm keV}}{\omega} \right) \left( \frac{{\rm cloud} \, {\rm size}}{10 \, {\rm pc}} \right) \gtrsim 2 \pi
\ee
In figure \ref{low} $\delta = 1.1$, whereas in figure \ref{high} $\delta = 5.5$. This condition is almost never satisfied in the Milky Way, so in this work we simply use the volume averaged electron density given in \cite{NE2001}. However, this effect is significant in high electron density environments such as starburst galaxies. Furthermore, an analogous effect will operate whenever oscillations are suppressed by an effective mass from astrophysical plasmas.

The warm neutral medium responsible for photoelectric absorption also has a cloud-like structure. We find that using a clumpy rather than homogeneous warm neutral medium for Milky Way densities only has a significant effect (after averaging over cloud positions) on $P_{a \to \gamma}$ for $\omega \lesssim 200 \, {\rm eV}$. At these energies, we found that photoelectric absorption reduces the expected signal to negligible levels in either case. We therefore simply use the volume averaged warm neutral medium densities.

\section{Starburst galaxies}

We now consider ALP to photon conversion in starburst galaxies. Starburst galaxies host strong magnetic fields of up to $\mathcal{O} \left( 100 \,  \mu {\rm G} \right)$ in the core regions with somewhat lower fields in the halo \cite{12095552,Beck:2012}, making them potentially very good ALP to photon converters. However, the fields in starbust galaxies are largely turbulent with little or no coherent field. Furthermore, the electron density is correspondingly higher at $\mathcal{O} \left( 100 - 1000  \, {\rm cm}^{-3} \right)$ \cite{09073162}. Naively, we might expect this high electron density to substantially suppress $P_{a \to \gamma}$. However, in this regime the cloud structure of the WIM becomes highly significant, as shown in section 5. The \emph{intercloud} electron density it also crucial here. For example, consider a simplified case with a random field $B = 150 \, \mu {\rm G}$ over a distance of 3 kpc and  coherent over 100 pc, implemented as described in section 3. We use a volume averaged electron density $\overline{n}_{e} = 1000 \, {\rm cm}^{-3}$. Using $\omega = 1 \, {\rm keV}$, $M = 5 \times 10^{12} \, {\rm GeV}$ and the constant, volume averaged electron density gives $P_{a \to \gamma} \sim 10^{-11}$, averaging over 100 instances of the random field. Now let us assume the WIM exists in 1 pc clouds (for example as in the starburst galaxy M82 \cite{09073162}) with filling factor $f = 0.1$. In this case, applying equation \eqref{delta} we obtain $\delta = 1100$ and so the presence of clouds is of great importance. If we assume an intercloud electron density of $0.1 \, {\rm cm}^{-3}$ we obtain $P_{a \to \gamma} \sim 10^{-4}$ - a conversion probability comparable to that in galaxy clusters. However, if we assume that the intercloud electron density is $10 \, {\rm cm}^{-3}$ we obtain $P_{a \to \gamma} \sim 10^{-7}$, rendering any signal unobservable (although still 4 orders of magnitude higher than for a volume averaged electron density). It is therefore possible, but certainly not guaranteed, that we might see signals from a CAB in some starburst galaxies. Furthermore, at ALP energies $E \lesssim 500 \, {\rm eV}$, any signal would by highly suppressed by photoelectric absorption. Starbursts might also be good observation targets for the 3.5 keV line arising from dark matter decay to ALPs discussed in \cite{14032370,14047741,14065518,14101867}.

\section{Conclusions}


We have simulated ALP to photon conversion probabilities for axion-like particles propagating through the Milky Way to Earth. We find that the cosmic axion background motivated by string models of the early universe and by the cluster soft X-ray excess would be entirely unobservable following ALP-photon conversion in the Milky Way's magnetic field. This is due to low conversion probabilities in the Milky Way relative to galaxy clusters, as well as the high photoelectric absorption cross section for the central CAB energies. Furthermore, the North-South asymmetry in this magnetic field is not reflected in observations of the unresolved cosmic X-ray background, ruling out a significant ALP contribution to the cosmic X-ray background. The smaller size of galaxies compared with galaxy clusters make them in general poorer targets for observation of ALP-photon conversion. One exception might be starburst galaxies, which feature very high magnetic fields and electron densities.

The galactic electron density suppresses conversion by giving an effective mass to the photon component, but in such high density environments the detailed local structure of the plasma must be considered. We have derived a condition for when the cloud structure of a galaxy's electron density is relevant for ALP-photon conversion. We find that in the Milky Way, and other typical spiral and elliptical galaxies, the cloud structure is not relevant. However, the cloud structure \emph{is} relevant in high electron density environments such as starburst galaxies. We found that when the cloud structure of the electron density is taken into account, the predicted ALP-photon conversion probability in starburst galaxies is increased by up to 8 orders of magnitude, depending on the assumed intercloud electron density.

\section*{Acknowledgements}

I thank Pedro Alvarez, James Bonifacio, Joseph Conlon, Edward Hughes, David Marsh, Andrew Powell and Markus Rummel for helpful discussions. I further thank Joseph Conlon for careful reading and advice on the manuscript and David Marsh for providing the CAB spectrum fit. This work is part funded by the European Re-search Council Starting Grant 307605-SUSYBREAKING. I am funded by an STFC studentship (ST/M50371X/1).

\appendix

\section{Comparison with previous results}

Our main conclusion - that a cosmic axion background cannot contribute significantly to the unresolved cosmic X-ray background - differs to that found by Fairbairn in \cite{Fairbairn} (hereafter F2013). Here we explain further why our results differ. The key elements are:

\begin{itemize}

\item Consistent conversion of electromagnetic quantities from natural units to SI units
\item Modeling of photoelectric absorption using the density matrix formalism
\item The ALP spectrum
\item The morphology of the unresolved X-ray background

\end{itemize}

We now consider each of these factors in turn.

\subsection{Conversion of $B$ from natural to SI units}

To compute conversion probabilities from equation \ref{propfree} we must convert $\Delta_{\gamma} = \frac{-\omega_{pl}^{2}} {2 \omega} =  -\frac{4 \pi \alpha n_{e}}{2 \omega m_{e}}$ and $\Delta_{\gamma i} = \frac{B_{i}}{2 M}$ from natural units to SI units. We will use natural Lorentz-Heaviside units such that $\alpha = \frac{e^{2}}{4 \pi} \simeq \frac{1}{137}$, with $\epsilon_{0} = \mu_{0} = 1$. To convert the magnetic field strength from natural Lorentz-Heaviside to SI units, we may consider for example the corresponding energy density in SI units:

\be 
\rho = \half \frac{B^{2}}{\mu_{0}}
\ee

We find that 1 Gauss ($=10^{-4} \, T$) corresponds to $1.95 \times 10^{-2} \, {\rm eV}^{2}$ in natural Lorentz-Heaviside units. (See also footnote 24 of Raffelt and Stodolsky \cite{Raffelt}).  Using this conversion factor, we find:

\begin{equation}
\label{mixingComparison} 
\Delta_{\gamma i} = \frac{B_{i}}{2 M} = 1.53 \times 10^{-5} \left( \frac{B_{i}}{\mu {\rm G}} \right) \left( \frac{10^{14} \, {\rm GeV}}{M} \right) \, {\rm kpc}^{-1}
\end{equation}

We now turn to the conversion of $\Delta_{\gamma}$. Using $\alpha = \frac{e^{2}}{4 \pi} \simeq \frac{1}{137}$ we obtain:

\begin{equation}
\label{massComparison}
\Delta_{\gamma} = \frac{- \omega_{pl}^{2}}{2 \omega} = 1.1 \left( \frac{n_{e}}{10^{-2} \, {\rm cm}^{-3}} \right) \left(\frac{{\rm keV}}{\omega} \right) \, {\rm kpc}^{-1}
\end{equation}

Comparing equations \ref{mixingComparison} and \ref{massComparison} to the corresponding expressions in F2013 (in between equations 9 and 10), we find that while our expressions for $\Delta_{\gamma}$ agree, F2013's $\Delta_{\gamma i}$ is a factor of $\sqrt{4 \pi}$ too high, resulting in conversion probabilities that are a factor $4 \pi$ too high.

\subsection{Treatment of photoelectric absorption}

Photoelectric absorption by the warm neutral medium is highly significant for photon energies in the ROSAT 1/4 keV band, but much less significant for the 3/4 keV band. We include this effect using the standard density matrix formalism (see for example \cite{DensityMatrix}) described in equations \ref{H} and \ref{discrete}. As shown in figures \ref{200Prob} and \ref{500Prob}, this leads to a substantial suppression of the ALP-photon conversion probability at $\omega = 200 \, {\rm eV}$ compared to that at $\omega = 500 \, {\rm eV}$  (around 2 orders of magnitude).

F2013 also considers photoelectric absorption, but uses a different propagation equation for the density matrix (see equations 10 and 11 in F2013). While he does not give his simulated conversion probabilities for energies in the 1/4 keV band, we can make some inferences from other information given. In section 2.2, F2013 finds that similar conversion probabilities are needed to explain the 1/4 keV band and 3/4 keV band signals. However, his figure 5 shows similar M values required to explain the 1/4 keV and 3/4 keV bands. From this we conclude that his calculated conversion probabilities for the 1/4 keV and 3/4 keV energies are similar, and thus the effects of photoelectric absorption in the 1/4 keV band have not been properly taken into account. 

\subsection{The ALP spectrum}

F2013 considers two ALP spectra:

\be
\frac{d \Phi_{357}}{d E} \simeq 64.5 \sqrt{\frac{\Delta N_{{\rm eff}}}{0.5} \frac{E}{{\rm eV}}} {\rm exp} \left[- \left( \frac{E}{357 \, {\rm eV}} \right)^{2} \right] \, {\rm cm}^{-2} {\rm s}^{-1} {\rm sr}^{-1} {\rm eV}^{-1}
\ee
(his equation 4), and

\be
\frac{d \Phi_{950}}{d E} \simeq 5.58 \sqrt{\frac{\Delta N_{{\rm eff}}}{0.5} \frac{E}{{\rm eV}}} {\rm exp} \left[- \left( \frac{E}{950 \, {\rm eV}} \right)^{2} \right] \, {\rm cm}^{-2} {\rm s}^{-1} {\rm sr}^{-1} {\rm eV}^{-1}
\ee
(his equation 5)

In his equation 2, F2013 (correctly) states that these spectra should be normalised such that:

\be 
\rho_{{\rm CAB}} = \frac{7}{8} \left(\frac{4}{11}\right)^{\frac{4}{3}} \Delta N_{{\rm eff}} \rho_{{\rm CMB}} = 5.9 \times 10^{-2} \Delta N_{{\rm eff}} \, {\rm eV} {\rm cm}^{-3}
\ee 

However, in fact both his spectra are normalised such that

\be 
\rho_{{\rm CAB}} = 2 \pi \frac{4}{c} \int_{0}^{\infty} \frac{d \Phi} {dE} E dE = 5.9 \times 10^{-2}  \, {\rm eV} {\rm cm}^{-3}
\ee

i.e. he has presumably failed to include the factor of $\Delta N_{{\rm eff}} = 0.5$.

Furthermore, F2013 states that the energy of ALP spectrum $\frac{d \Phi_{950}}{d E}$ in the 0.64-1 keV band is $9.2 \times 10^{7} \, {\rm eV} \, {\rm s}^{-1} {\rm cm}^{-2} {\rm sr}^{-1}$ (the reported units are actually ${\rm eV}^{-1}$, but I assume this is a typo). However, using F2013's own spectrum, the energy is:

\be
\int_{0.64 \, {\rm keV}}^{1 \, {\rm keV}} \frac{d \Phi_{950}} {dE} E dE = 2.2 \times 10^{7} \, {\rm eV} \, {\rm s}^{-1} {\rm cm}^{-2} {\rm sr}^{-1}
\ee

Similarly, F2013 states that the energy of ALP spectrum $\frac{d \Phi_{357}}{d E}$ in the 100-284 eV band is $9.3 \times 10^{7} \, {\rm eV} \, {\rm s}^{-1} {\rm cm}^{-2} {\rm sr}^{-1}$.  However, integrating the spectrum he gives in equation 4 we find

\be
\int_{0.1 \, {\rm keV}}^{0.284 \, {\rm keV}} \frac{d \Phi_{357}} {dE} E dE = 2.3 \times 10^{7} \, {\rm eV} \, {\rm s}^{-1} {\rm cm}^{-2} {\rm sr}^{-1}
\ee

These two mistakes combined mean that F2013 has overestimated the ALP flux itself by a factor of almost 10. We also note that F2013 uses an ALP spectrum with a higher average energy when attempting to reproduce the 3/4 keV band flux. Since the publication of F2013, CAB spectra with average ALP energies $E_{{\rm CAB}} \gtrsim  250 \, {\rm eV}$ were excluded by overproduction of X-rays in galaxy clusters \cite{13123947}. We therefore do not consider such models here. 

\subsection{Morphology}

F2013's prediction of an observable X-ray signal from CAB to photon conversion in the Milky Way magnetic field is due to the compounding errors detailed above. We also note here that, as discussed in section \ref{CAB}, the morphologies of the Milky Way magnetic field and the unresolved X-ray background make a significant CAB contribution to the unresolved X-ray background impossible. This issue is not considered in F2013.

\bibliography{refs}
\bibliographystyle{elsarticle-num}

\end{document}